\documentclass[epj]{svjour}
\usepackage{color}
\usepackage{amsmath}
\usepackage{graphicx}
\usepackage{dcolumn}
\usepackage{bm}
\usepackage{hyperref}
\usepackage{longtable}
\usepackage{supertabular}
\usepackage{latexsym}
\usepackage{graphics}
\usepackage{epstopdf}
\begin{document}
\title{Tensor force effect on the neutron shell closure in super-heavy elements}
\author{M. EL ADRI\inst{1}\thanks{\emph{corresponding author:}  mohamed.eladri@ced.uca.ma}\and M. Oulne\inst{1}
%
}                     
%
%
\institute{High Energy Physics and Astrophysics Laboratory, Department of Physics,  Faculty of Sciences Semlalia, Cadi Ayyad University,
P.O.B. 2390, Marrakesh 40000, Morocco.}
\date{Received: date / Revised version: date}
%
\abstract{
A systematic study of the effect of tensor force on the evolution of shell structure in even-even super-heavy nuclei in the region of proton numbers Z=114, 120 and 126 and in the region of neutron numbers 178 $\le$ N $\le$ 188 is presented. We use, in this investigation, the Hartree-Fock framework by means of different types of Skyrme functionals in two cases with and without tensor force. The Bardeen-Cooper-Schriefer (BCS) approximation has been used to treat the pairing correlations. By investigating structural and decay properties of nuclei under consideration, it is found that N=184 shell gap is more enhanced by the tensor interaction which depends on the isoscalar tensor coupling constant $C_0^J$ of the used Skyrme interactions.  In the case without tensor interaction, this gap is significant only for T22, T24, T42 and SLy5. So, it  disappears with T46, T64 and T66, and is too weak for T26, T44 and T62. Without exception, the shell gap at N=184 becomes more pronounced when the tensor part is taken into account.
} 

\maketitle
 \section{introduction}
Recently, due to the progress in experimental techniques, the scientists have synthesized six new super-heavy elements with proton numbers Z =113 (Nh, nihonium), Z=114 (Fl, flerovium) \cite{Barber2011}, Z=115 (Mc, moscovium), Z=116 (Lv, livermorium), Z=117 (Ts, tennessine), and Z=118 (Og, oganesson) \cite{Karol2016}. This discovery refreshes the research in this region and triggers anew the curiosity of nuclear physicists. Even that, predicting a probable island of stability of super-heavy nuclei remains a major challenge. Till now, a really next doubly magic nuclei beyond Pb (Z=82, N=126) is not yet known. The microscopic-macroscopic models  \cite{Nilsson,Moller} predict (Z = 114, N = 184) as a doubly magic nuclei. However, the most stable nuclei in relativistic \cite{Rutz,Zhang} and non-relativistic \cite{Cwiok,Bender1999} density functional theories (DFTs) are (Z=120, N=172/184) and (Z=126, N=184), respectively. So far, the center of the island of stability in the SHN region, predicted by these approaches, has not yet been reached experimentally despite recent attempts to synthesize isotopes of the element Z=120 \cite{Oganessian2006,Oganessian2009}.

It is worth noting that the shell structure of super-heavy nuclei studied within DFT framework is most accurate then that studied within micro-macroscopic models. This is due to the lack of self-consistency effect in the micro-macroscopic models with respect to DFT framework \cite{Agbemava2015}. This motivates us to study the shell structure of SHN using the Energy Density Functionals (EDFs).
On the other hand, since the introduction of the tensor force, in 1940, by Bethe \cite{Bethe1940a,Bethe1940b}, and investigating its effect on the deuteron property, the physicists have treated it as a necessary element in the nucleon-nucleon (NN) interaction. %
Historically speaking, Skyrme formally included the tensor force into his interaction as shown in \cite{Skyrme1959-1,Skyrme1959-2,Skyrme1959-3}. Stancu and Brink \cite{Stancu} discussed tensor-force effects in the famous Vautherin-Brink model of Skyrme interaction \cite{Vautherin}.
However, in enormous amounts of the works by this model, the tensor force had been overlooked for three decades, a surprisingly long period. Only particularly after \cite{Otsuka2001} in 2001, at first time, and \cite{Otsuka2005} in 2005, the significance of the tensor force was realized by many authors, including the inspiration by the first attempt to DFT, realized by Otsuka et al. \cite{Otsuka2006}, after Stancu et al. \cite{Stancu}.
After that the range of the interactions fitted with the tensor part became more large.
Currently, the mean-field theories allow to include the tensor force in standard Skyrme interactions, such as the Skx \cite{Brown2006}, SLy5+T \cite{Colo} SLy4+T, SLy4T$_{min}$\cite{Zalewski}, SLy4T$_{self}$ and TZA \cite{Bender}. Furthermore, Lesinski et al. built a new set of 36 parametrizations (TIJ) \cite{Lesinski}, covering a wide range of the parameter space of the isoscalar and isovector tensor term coupling constants with a fit protocol very similar to that of the successful SLy parametrizations. By means of these functionals, many works have shown that the tensor force plays a crucial role in the investigation of the ground state features of nuclei, whereas Otsuka and collaborators were the first to study the influence of this force on the single-particle energies as well as on the spin-orbit interactions for exotic nuclei \cite{Otsuka2001,Otsuka2005}. The standard Skyrme Hartree-Fock calculations cannot describe the experimental isospin dependence of the spin-orbit splitting, whereas it is well reproduced by including the interaction of the tensor force. 

In superheavy nuclei, due to high level density of single-particle energies, a small variation in the single particle energies may significantly affect shell structure. This variation is  responsible for the different predictions of magic numbers in SHN region. As pointed out in the previous paragraph, the tensor force plays a crucial role in describing the shell gaps. Therefore, it is necessary to take into account the effect of this force in such study. In last decade, the influence of the tensor component on the shell closure of SHN in the Hartree-Fock framework as well as in the Skyrme Hartree–Fock+BCS model was investigated \cite{Suckling,Zhou}. In our present work, similar study has been done ch a large variety of new Skyrme sets have been used to investigate the SHN lying in the island of stability and in their vicinity covering the region of proton numbers Z=114, 120 and 126 and the region of neutron numbers 178 $\le$ N $\le$ 188. 

The article is organized as follows: Sect \ref{Sec2} shortly summarizes the approaches that we have used to do our calculations. In sect \ref{Sec3} numerical tests as well as the input details and the interactions used in calculations are presented. The obtained results are analyzed and discussed in Sect \ref{Sec4}. Finally, the main conclusions and outlook are given in Sect \ref{Sec5}. 

\section{Theoretical framework} \label{Sec2}
The Skyrme Hartree-Fock plus Bardeen-Cooper-Schrieffer (HF+BCS) theory \cite{Ryssens2015} with effective zero-range pairing interactions is a practical computational tool to investigate the nuclear pairing correlations in nuclei near to $\beta$-stability. As the HF+BCS approach has been widely discussed in the literature \cite{Ryssens2015,Bender}, it will be briefly presented here.\\
\subsection{The standard Skyrme force}
The standard Skyrme force is composed of three components, namely: the central, tensor, and spin-orbit interactions
\begin{equation}
\hat{\nu}=\hat{\nu}^{central}+\hat{\nu}^{tensor}+\hat{\nu}^{LS}
\end{equation}
The central two-body Skyrme interaction can be written as \cite{Vautherin}
\begin{eqnarray}
\hat{v}^{central} (\mathbf{r,r'}) & & = t_0 \, ( 1 + x_0 \hat{P}_\sigma ) \; \delta (\mathbf{r-r'})
\nonumber \\
&   & + \frac{1}{2} \, t_1 \, ( 1 + x_1 \hat{P}_\sigma ) \,
\Big[\mathbf{\hat{k}}^{\prime 2} \; \delta (\mathbf{r-r'})
+ \delta (\mathbf{r-r'}) \; \mathbf{\hat{k}}^2
\Big]
\nonumber \\
&   & + t_2 \ ( 1 + x_2 \hat{P}_\sigma ) \,
\mathbf{\hat{k}}^{\prime} \cdot \delta (\mathbf{r-r'}) \; \mathbf{\hat{k}}
\nonumber \\
&   & + \frac{1}{6} \, t_{3a} \, ( 1 + x_{3a} \hat{P}_\sigma ) \;
\rho_0^{{\alpha_a}} (\mathbf{R}) \; \delta (\mathbf{r-r'})
\nonumber \\
&   & + \frac{1}{6} \, t_{3b} \, ( 1 + x_{3b} \hat{P}_\sigma ) \;
\rho_0^{\alpha_b} (\mathbf{R}) \; \delta (\mathbf{r-r'}) \, ,
\end{eqnarray}\\
In the above expression, the quantity $\hat{P}_\sigma$ represents the spin exchange operator, the operator
$\hat{\mathbf{k}} = - \frac{\mathrm{i}}{2}(\nabla -\nabla')$ is the relative momentum operator which acts on the right and $\hat{\mathbf{k}'}$ is its complex conjugate acting to the left, and $\rho_0 (\mathbf{R})$ is the isoscalar nucleonic density at $\mathbf{R} = \frac{1}{2} ( \mathbf{r} + \mathbf{r}' )$.\\
The spin-orbit component is defined by \cite{Bell1972,Skyrme1958}
\begin{eqnarray}
\hat{v}^{LS} (\mathbf{r,r'}) =  i W_0 \, ( \hat{\mathbf{\sigma}}_1 + \hat{\mathbf{\sigma}}_2 ) \cdot
\hat{\mathbf{k}}^{\prime} \times
\delta (\mathbf{r,r'}) \;  \hat{\mathbf{k}}
\end{eqnarray}
The Skyrme two zero-range tensor term reads \cite{Skyrme1956}
\begin{eqnarray}
\hat{v}^{tensor} (\mathbf{r,r'}) & & =  \frac{1}{2} \; t_e \; \Big\{ \Big[ 3 \,( \mathbf{\sigma_1} \cdot \mathbf{k}' ) \, ( \sigma_2 \cdot \mathbf{k}' )
- ( \sigma_1 \cdot \sigma_2 ) \, \mathbf{k}^{\prime 2} \,
\Big] \; \delta (\mathbf{r-r'})
\nonumber \\
& &  
+  \delta (\mathbf{r-r'}) \;
\Big[ 3 \, ( \sigma_1 \cdot \mathbf{k} ) \, ( \sigma_2 \cdot \mathbf{k} )
-  ( \sigma_1 \cdot \sigma_2) \, \mathbf{k}^{2}
\Big]
\Big\}
\nonumber \\
&  & + \frac{1}{2} t_o \,
\Big\{\Big[
3 \, ( \sigma_1 \cdot \mathbf{k}' ) \, \delta (\mathbf{r-r'})\,
( \sigma_2 \cdot \mathbf{k} )
\nonumber \\
& &  
-  ( \sigma_1 \cdot \sigma_2 ) \, \mathbf{k}' \cdot \,
\delta (\mathbf{r-r'}) \, \mathbf{k}
\Big]
\nonumber \\
&  &
+ \Big[ 3 \, ( \sigma_2 \cdot \mathbf{k}' ) \, \delta (\mathbf{r-r'})\,
( \sigma_1 \cdot \mathbf{k} )
\nonumber \\
&  & 
-  ( \sigma_1 \cdot \sigma_2 ) \, \mathbf{k} \cdot \,
\delta (\mathbf{r-r'}) \, \mathbf{k}'
\Big]\Big\}
\, .
\end{eqnarray}
This expression contains two tensor interactions: An “even” tensor force with the coupling constant $t_e$ and an “odd” tensor force with the coupling constant $t_0$. It contains also the vectors formed by the Pauli spin matrices  $\sigma_1$ and $\sigma_2$.
The tensor interactions contribute to the binding energy and spin-orbit splitting. with this contribution, the SHF energy density is given by \cite{Stancu}
\begin{eqnarray}
\mathcal{H}^t=\frac{1}{2}\alpha(\mathbf{J_n}^2+\mathbf{J_p}^2)+\beta\mathbf{J_n}\cdot\mathbf{J_p}^2
\end{eqnarray}
where 
\begin{eqnarray}
\alpha\quad=C_0^J+C_1^J, \qquad \beta \quad=C_0^J-C_1^J\\ \nonumber
C_0^J= \frac{1}{2}(\alpha+\beta), \qquad C_1^J= \frac{1}{2}(\alpha-\beta)
\end{eqnarray}
The proton-neutron coupling constants $\alpha= \alpha_C+\alpha_T$ and $ \beta=\beta_C+\beta_T$ can also be separated into contributions from central (eq. \ref{Eq:C}) and tensor forces (eq. \ref{Eq:T}):\\
\begin{eqnarray}\label{Eq:C}
\alpha_C&&=\frac{1}{2}(t_1-t_2)-\frac{1}{2}(t_1x_1-t_2x_2), \\\nonumber
\beta_C&&=\frac{1}{2}(t_1x_1-t_2x_2)\\\nonumber
\alpha_T&&=\frac{5}{4}t_0, \\\nonumber
\beta_T&&=\frac{5}{8}(t_e+t_0) \label{Eq:T}
\end{eqnarray}
The spin-orbit potential,  with the contributions of tensor force correction, is written as
\begin{eqnarray}
W_q(r)=\frac{W_0}{2}(2\nabla\rho_q+\nabla\rho_{q'})+\alpha J_q+\beta J_{q'}
\end{eqnarray}
Here, q (q') represents the interactions between like (unlike) particles. The $J_{q(q')}(\mathbf{r})$ is the proton or neutron spin current density defined as
\begin{eqnarray}
J_{q(q')}(\mathbf{r})& &=\frac{1}{4 \pi r^3} \sum_i (2j_i+1)\\\nonumber
& & \times \Big[j(j+1)-l(l+1)-\frac{3}{4}\Big]R_i^2(r)
\end{eqnarray}
where i = n, l, j runs over all states having the given q(q'), and $R_i(r)$ is the radial part of the wave function.

\subsection{The energy density functional}
In the HF+BCS framework with the Skyrme interaction, the total binding energy is given by the sum of the kinetic energy ($E_{kin}$), the Skyrme energy ($E_{Sky}$) that describes the effective interaction between nucleons, the Coulomb energy ($E_{Coul}$), the pairing energy ($E_{pair}$), and a center-of-mass ($E_{c.m}$) correction
\begin{equation}
E = E_{kin} + E_{Sky} + E_{Coul} + E_{pair} + E_{c.m}
\end{equation}
The kinetic energy defined as
\begin{equation}
E_{kin}=\sum_{q=n,p}\int d^3r \frac{\bar{h}}{2m_q}\tau_q(r)
\end{equation}
The pairing energy contribution to the total energy is
\begin{equation}
E_{pair}=\sum_{k,m>0}f_k u_k \nu_k f_m u_m \nu_m \bar{\nu}^{pair}_{k\bar{k}m\bar{m}}
\end{equation}
Where the $\bar{\nu}^{pair}_{k\bar{k}m\bar{m}}$ are antisymmetrized matrix  elements of the pairing interaction and the $f_i$ are cutoff factors that are, both specified in section \ref{Sec3}.\\
The Skyrme energy density functional is written in the form
\begin{eqnarray}
E_{Sky} & &=\sum_{t=0,1}\int d^3r\Big(C^{\rho}_t[\rho_0]\rho_t^2+C^{\rho^{\alpha}}_t\rho_0^{\alpha}\rho_t^2+C^{\tau}_t\rho_t\tau_t\\\nonumber
&&+C^{\Delta\rho}_t\rho_t\Delta\rho_t
+C^{\nabla\rho_J}_t\rho_t\nabla.\mathbf{J}_t-C^T_t\sum_{\mu\nu}J_{t,\mu\nu}J_{t,\mu\nu}\Big)
\end{eqnarray}
This expression consists of the contribution of central, spin–orbit and tensor terms to the energy density functional.\\
The coupling constants are defined in Ref \cite{Ryssens2015}. 
The Coulomb energy $E_{Coul}$ contains two terms. The first one is a direct term that is dependent on the local density
\begin{equation}
E_{Coul} ^{dir}=\frac{e^2}{2}\int \int d^3r d^3r' \frac{\rho_p(r) \rho_p(r')}{|r-r'|}
\end{equation}
and the second one is an exchange term  that is calculated in the very efficient local Slater approximation \cite{Bender2003}
\begin{equation}
E_{Coul} ^{ex}=-\frac{3e^2}{4} \Big(\frac{3}{\pi}\Big)^{1/3}\int d^3r \rho_p^{4/3}(r) 
\end{equation}
The contribution of center-of-mass correction to the total energy can be estimated to be
\begin{equation}
E_{c.m}=\frac{(\sum_{i}^{A}P_i)^2}{2mA}
\end{equation}

\section{Numerical Details}\label{Sec3}
We use the EV8 code \cite{Ryssens2015} to perform HF+BCS calculations. This code solves the nuclear Skyrme -Hartree-Fock+BCS problem by discretizing the single-particle wave-functions on a three-dimensional Cartesian mesh and the imaginary time step method.\\
Our goal is to predict the influence of the tensor force on shell structure. For this purpose we have used a large variety of Skyrme parametrizations, namely: SLy5+T \cite{Colo}, T22, T24, T26, T42, T44, T46, T62, T64 and T66 \cite{Lesinski}. For each parametrization we treat two cases with and without tensor force. In SLy5+T, the tensor interactions are included to the existing set SLy5 to describe the spin-orbit splitting of Sn (Z=50) isotopes and N=82 isotones \cite{Colo}. However, the set of parametrizations T22, T24, T26, T42, T44, T46, T62, T64 and T66 are built with a fit protocol, similar to that of the SLyx parametrizations, of all parameters introducing tensor component to reproduce various empirical properties of nuclei along the chart nuclei \cite{Lesinski}. They have been successfully employed to study the evolution of shells for spherical \cite{Lesinski} and deformed \cite{Bender} nuclei, and also to analyze high spin states \cite{Hellemans}. Table \ref{table} lists the values of parameters $\alpha_T$, $\alpha_C$, $\beta_T$, $\beta_C$, the isoscalar tensor coupling constant $C_0^J$ and the isovector tensor coupling constant $C_1^J$ corresponding to different TIJ interactions as well as SLy5+T.

\begin{table*}[ht]
	\caption{The values of parameters $\alpha_T$, $\alpha_C$, $\beta_T$, $\beta_C$, the isoscalar tensor coupling constant $C_0^J$ and the isovector tensor coupling constant $C_1^J$ corresponding to different TIJ interactions as well as SLy5+T. All	values are in units of $MeV.fm^5$.\label{table}}
	\begin{center}
		{\begin{tabular}{@{}cccccc@{\hspace{18pt}}@{}c@{}} \hline
				\hline Parametrizations & $\alpha_T$ &  $\beta_T$ & $\alpha_C$ & $\beta_C$ & $ C_0^J$ & $C_1^J$\\ 
				\hline T22 & -90,630 & 28,862 & 807,882 & -329,908 & 0 & 0 \\ 
				\hline T24 & 24,674 & 19,365 & 780,285 & -321,520 & 60 & 60 \\ 
				\hline T26 & 150,873 & 31,758 & 821,1107 & -337,570 & 120 & 120 \\ 
				\hline T42 & -122,024 & 91,214 & 646,497 & -273,544 & 60 & -60 \\ 
				\hline T44 & 8,967 & 113,022 & 715,575 & -299,357 & 120 & 0 \\
				\hline T46 & 131,091 & 117,548 & 732,690 & -306,861 & 180 & 60 \\ 
				\hline T62 & -130,802 & 196,367 & 608,386 & -259,708 & 120 & -120 \\ 
				\hline T64 & -0,246 & 217,958 & 678,677 & -285,861 & 180 & 0 \\ 
				\hline T66 & 112,893 & 204,053 & 637,843 & -273,203 & 240 & 0 \\ 
				\hline SLy5+T & -170 & 100 & 870,198 & -353,933 & -19.333 & -70.466  \\  \hline
		\end{tabular}}
	\end{center}
\end{table*}

In the particle-hole channel, the density-dependent pairing interaction is given by \cite{Rigollet}
\begin{equation}	
\hat{\nu}^{pair}(\mathbf{r},\mathbf{r'})=\frac{V_0}{2}(1-\hat{P}_{\sigma})\left[1-\alpha\left(\frac{\rho_0(\mathbf{R})}{\rho_s}\right)\right]\delta(\mathbf{r}-\mathbf{r'})
\end{equation}
Here $\rho_0(\mathbf{R})$ is the isoscalar nucleonic density at $\mathbf{R}=\frac{1}{2}(\mathbf{r}+\mathbf{r'})$. The parameters are fixed at  $V_0=-1250MeV.fm^{-3} $, $\rho_s=0.16fm^{-3}$ and $\alpha=0.5$ called mixed volume-surface pairing. In order to avoid a basis-size dependence of the total energy which can  lead to non-physical divergences, two cuttofs of the pairing are introduced: one above and the other below the Fermi energy. They are defined by the product of two Fermi functions
\begin{equation}
f_k=[1+e^{(\epsilon_k-\lambda_q-\Delta\epsilon_q)/\mu_q}]^{-1/2}[1+e^{(\epsilon_k-\lambda_q+\Delta\epsilon_q)/\mu_q}]^{-1/2}
\end{equation}
where $\epsilon_k$ is the single-particle energy of the single-particle state k and $\lambda_k$ is the Fermi energy. Here, the $ \mu_k$ is fixed to 0.5MeV and the $\Delta\epsilon_q$ is set to 5MeV for both neutrons and protons.\\
The accuracy of a coordinate-space calculation is limited
by the size of the box $N_{Box}$, the discretization length $dx$ and the iterations number $N_{iter}$ \cite{Ryssens2015-2}. In this work, we have performed all calculations on a large box size $N_{Box}=26fm^3$ with a step size of $1.0fm$ between discretization points. The numerical time-step is set to $\delta t=0,012.10^{-22}s$ and the iterations number is fixed at $N_{iter}=500$.\\
\section{Results and Discussion}\label{Sec4}
In this section, we present results for the istopoic chain of even-even super-heavy nuclei with Z=114, Z=120 and Z=126 protons. Calculating several physical quantities such as two-neutron separation energies, two-neutron shell gap, neutron pairing energy, neutron single particle energies and $Q_{\alpha}$-decay properties, we discuss the impact of the tensor force on the neutron shell closure N=184 by means of HF+BCS approach using the ten set of Skyrme parametrizations mentioned in the previous section. The obtained results are compared with the available predictions of FRDM \cite{Moller2016}.
\subsection{Two-neutron separation energies }\label{S2nSec}
\begin{figure*}[htp]
	\begin{minipage}{0.5\textwidth}
		\centering \includegraphics[scale=0.3]{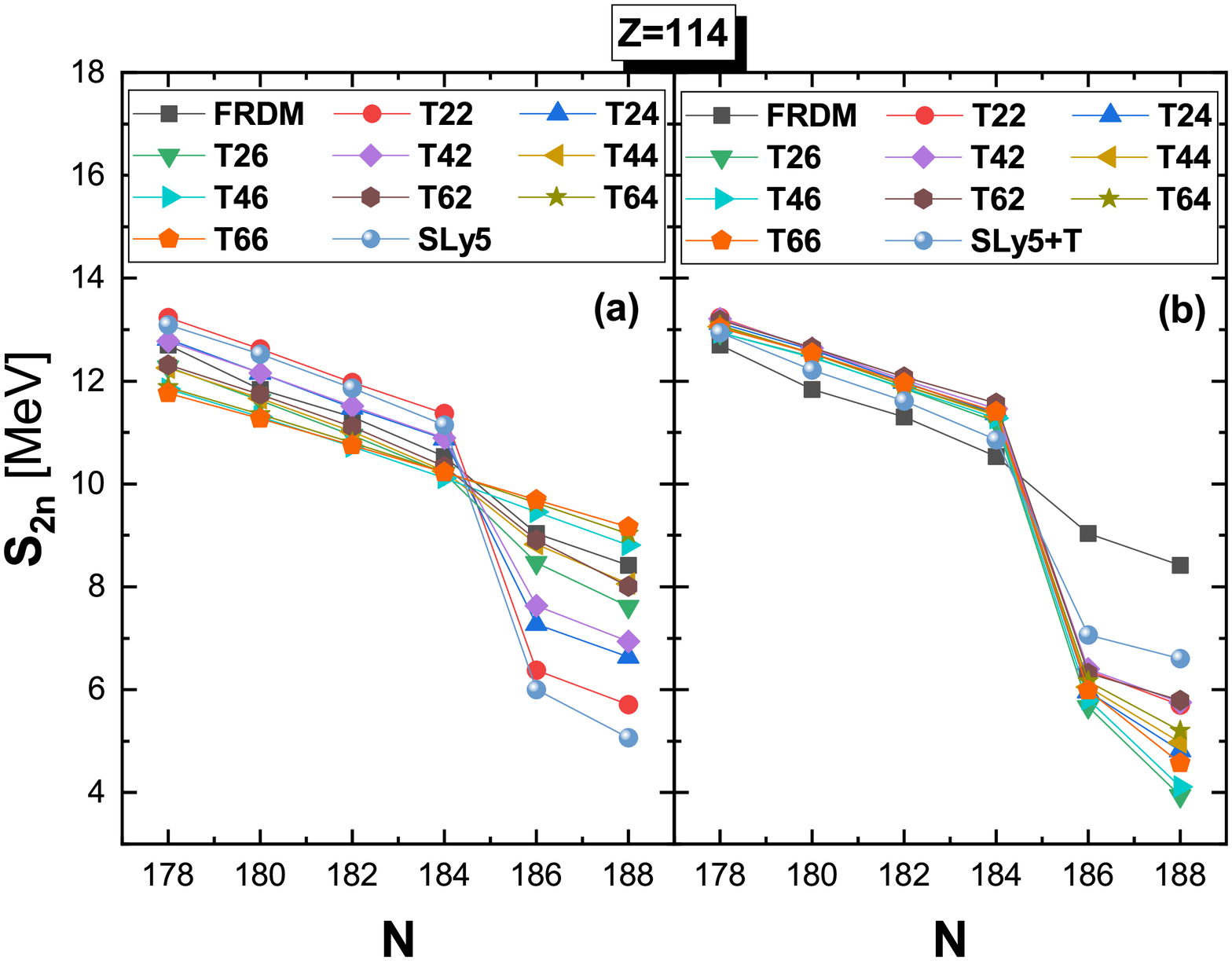}
	\end{minipage}
	\begin{minipage}{0.5\textwidth}
		\centering \includegraphics[scale=0.3]{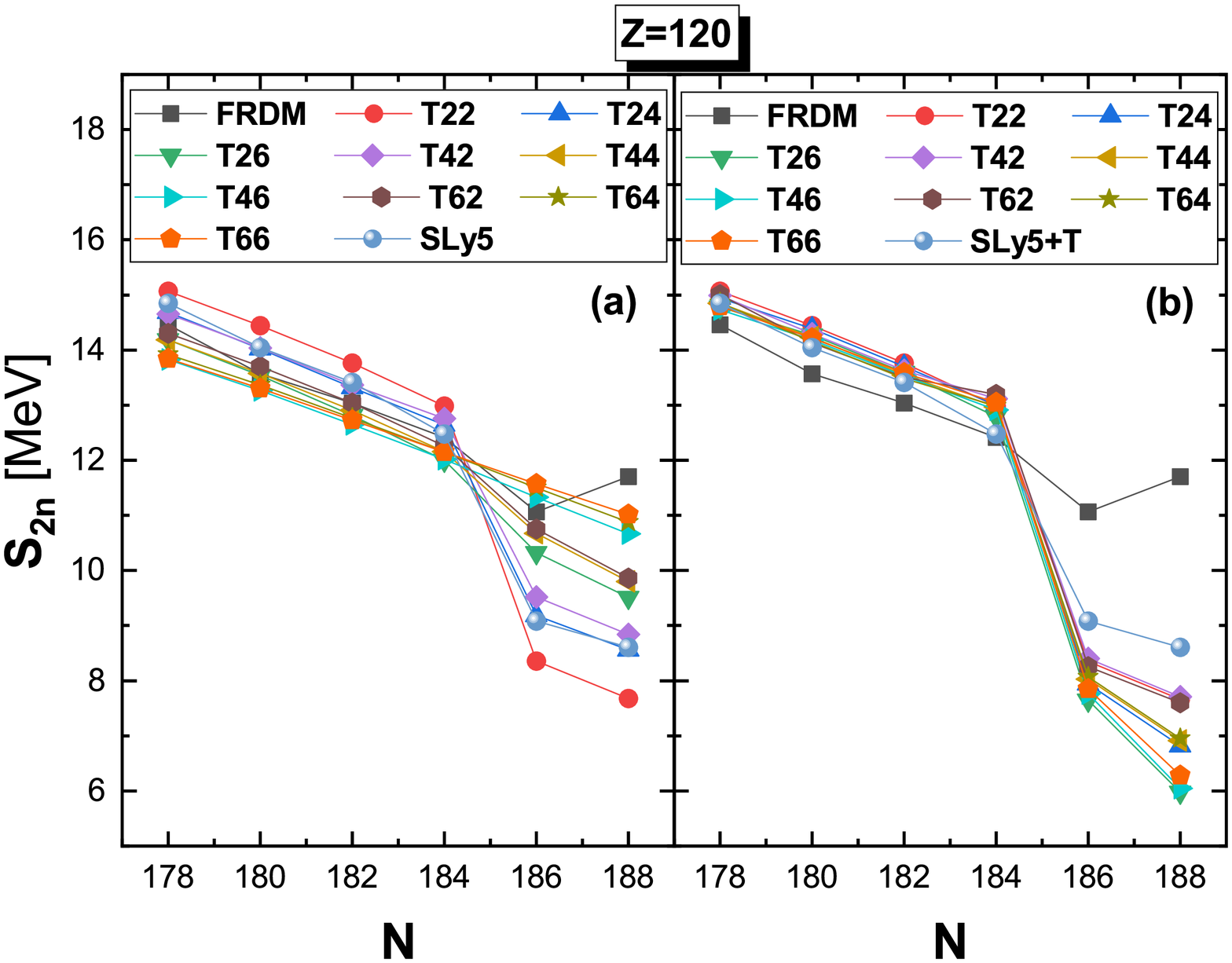}
	\end{minipage}\hfill
	\begin{minipage}{0.5\textwidth}
		\centering \includegraphics[scale=0.3]{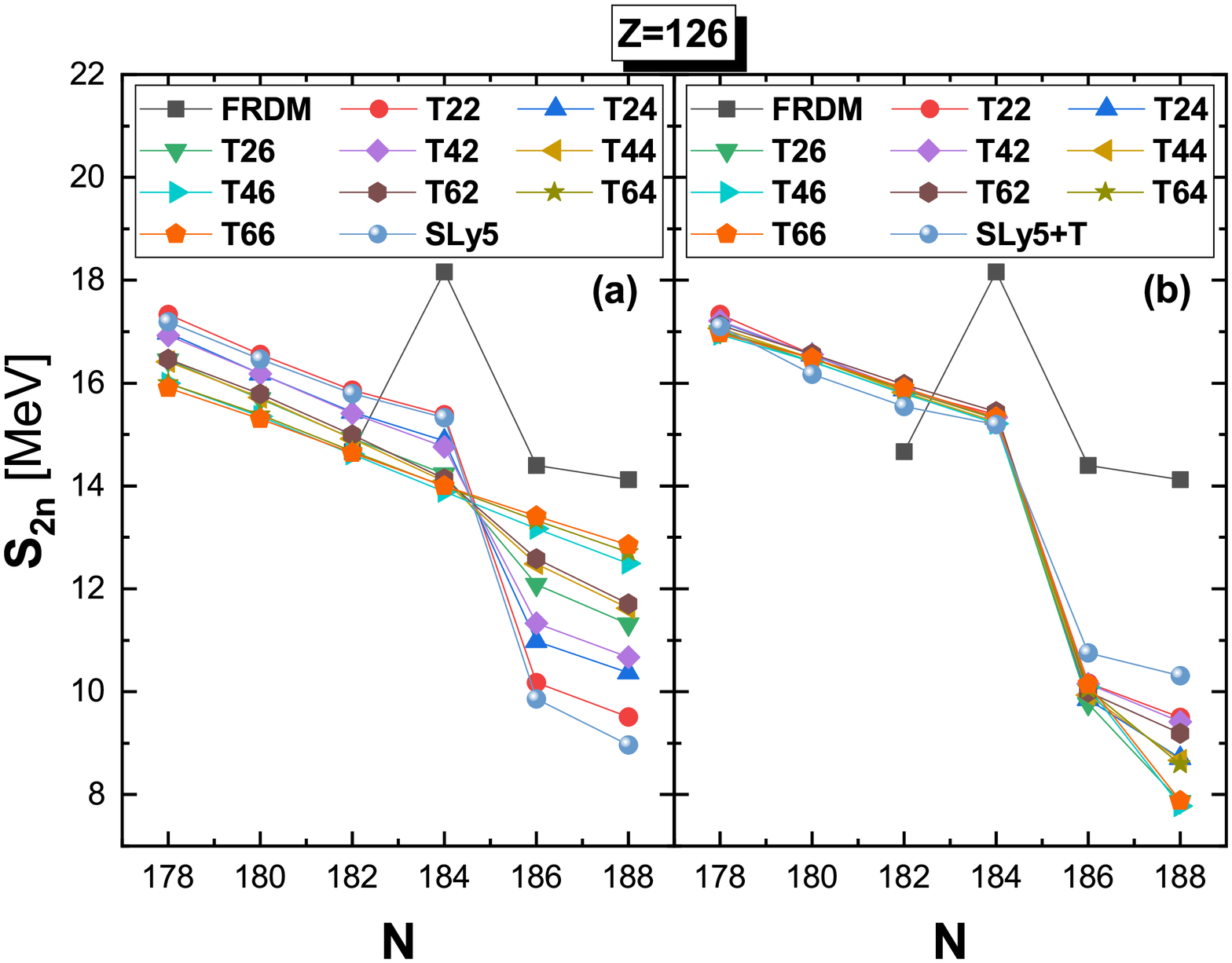}
	\end{minipage}\hfill
	\caption{Two-neutron separation energies $S_{2n}(Z,N)$ given for different isotopic chains under investigation as a function of neutron number N. The panels (a) show the calculated results without tensor force and the panels (b) illustrate the calculated results with tensor force. We compare our finding with FRDM predictions \label{S2nFig}}
\end{figure*}

Two-neutron separation energy of a nucleus with $Z$ protons and $N$ neutrons denoted as $S_{2n}(Z,N)$, is an essential physical property of nuclei which allows one to identify the shell closures. It is expressed as a function of binding energy $BE$ as follows
\begin{equation} 
S_{2n}(Z,N)=BE(Z,N)-BE(Z,N-2)
\end{equation}
In Fig. \ref{S2nFig} we plot our results of two-neutron separation energy $S_{2n}$ with the neutron number $N$ for the isotopic chains of  SHN with Z=114, Z=120 and Z=126 in the cases with and without the tensor force. Our findings are compared to FRDM predictions \cite{Moller}. For the SLy5, T22, T24 and T42 interactions, the quantity $S_{2n}$ presents a sharp drop at the closure shell N=184 for all three chains in both cases. This shell is less pronounced for T26, T44 and T62. However, for the sets T46, T64 and T66 there is only a jump at this neutron number when the contribution of the tensor force is present. This is because of lower isoscalar tensor coupling constant $C_0^J$ for SLy5, T22, T24 and T42 interactions, i.e. $C_0^J=0$, $C_0^J=-19.33MeV.fm^5$, $C_0^J=60MeV.fm^5$ and $C_0^J=60MeV.fm^5$, respectively, compared to $C_0^J=120MeV.fm^5$ for T26, T44 and T62, and  $C_0^J=240MeV.fm^5$ for T66. The T46 and T64 functionals share the same value of the isoscalar tensor coupling constant $C_0^J=180MeV.fm^5$. This implies that the tensor force effects on the gap N=184 become more and more evident for larger coupling constant $C_0^J$.\\
We can also see from the Fig. \ref{S2nFig} that for selected parametrizations, the values of the quantity  $S_{2n}$ without tensor force are lower than those with this force. In this latter case, the values of  $S_{2n}$ are almost the same for all Skyrme parametrizations, except SLy5+T which exhibit a slight difference in order of 0.3MeV compared to TIJ sets. So, the quantity $S_{2n}$ is enhanced due to the tensor force effects especially from the interactions with large $C_0^J$. Furthermore, the mac-microscopic model (FRDM) has less enhanced gap at N=184 for the isotopic series of Z=114 and Z=120 with respect to the microscopic ones. But, for Z=126 isotopes, FRDM predictions show an unexplained variation of $S_{2n}$. As we know, this amount, usually, evolves in a decreasing way as we move towards the higher mass region and shows a sharp jump at eventual shell closure. However, here, an important peak is observed at N=184, which indicates that the FRDM model fail to describe the N=184 shell closure in Z=126 nucleus. As pointed out in the introduction, this can be justify by the lack of self-consistency effect in the FRDM model with respect to DFT one.
\subsection{Two-neutron shell gap}
\begin{figure*}[htp]
	\begin{minipage}{0.5\textwidth}
		\centering \includegraphics[scale=0.3]{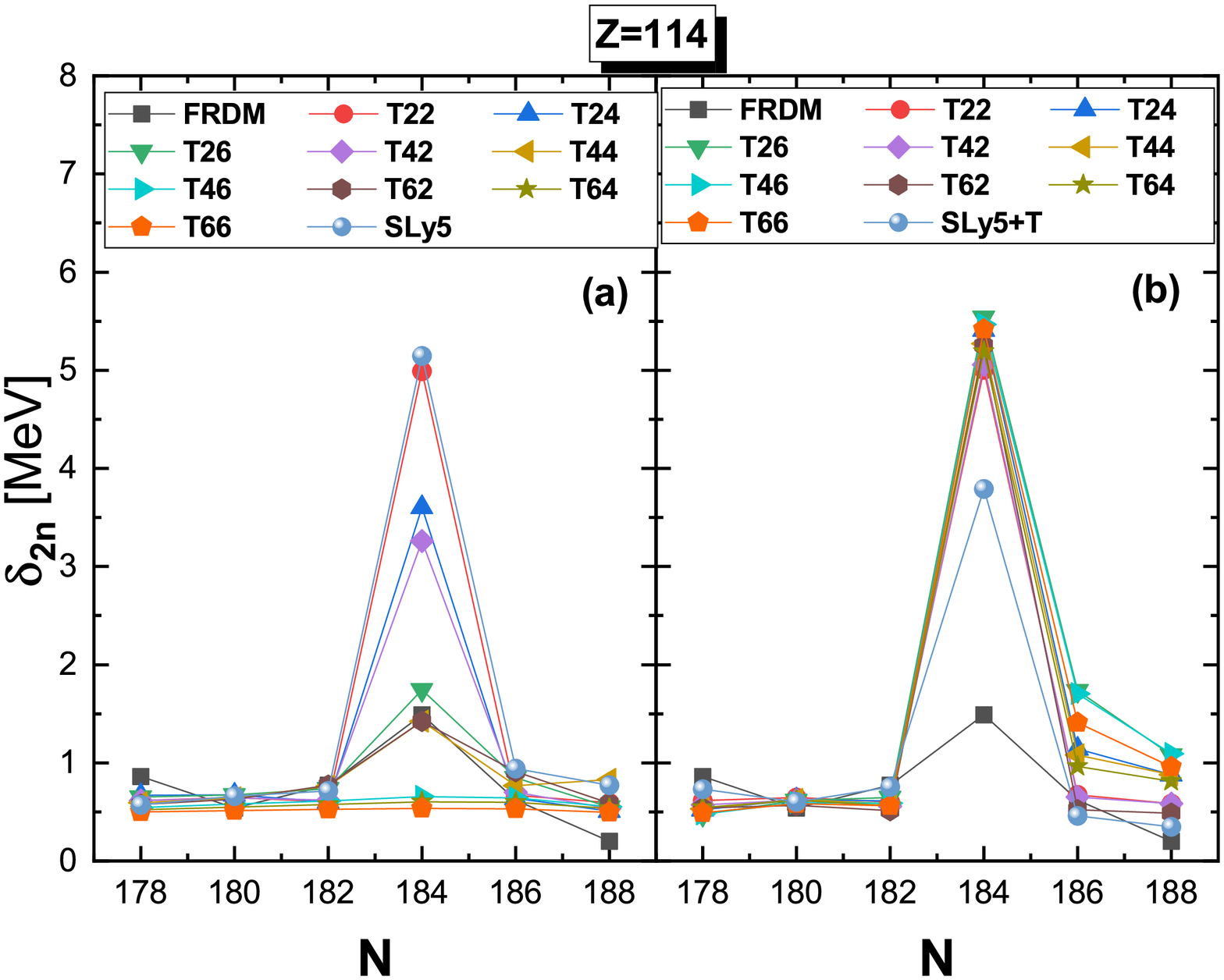}
	\end{minipage}
	\begin{minipage}{0.5\textwidth}
		\centering \includegraphics[scale=0.3]{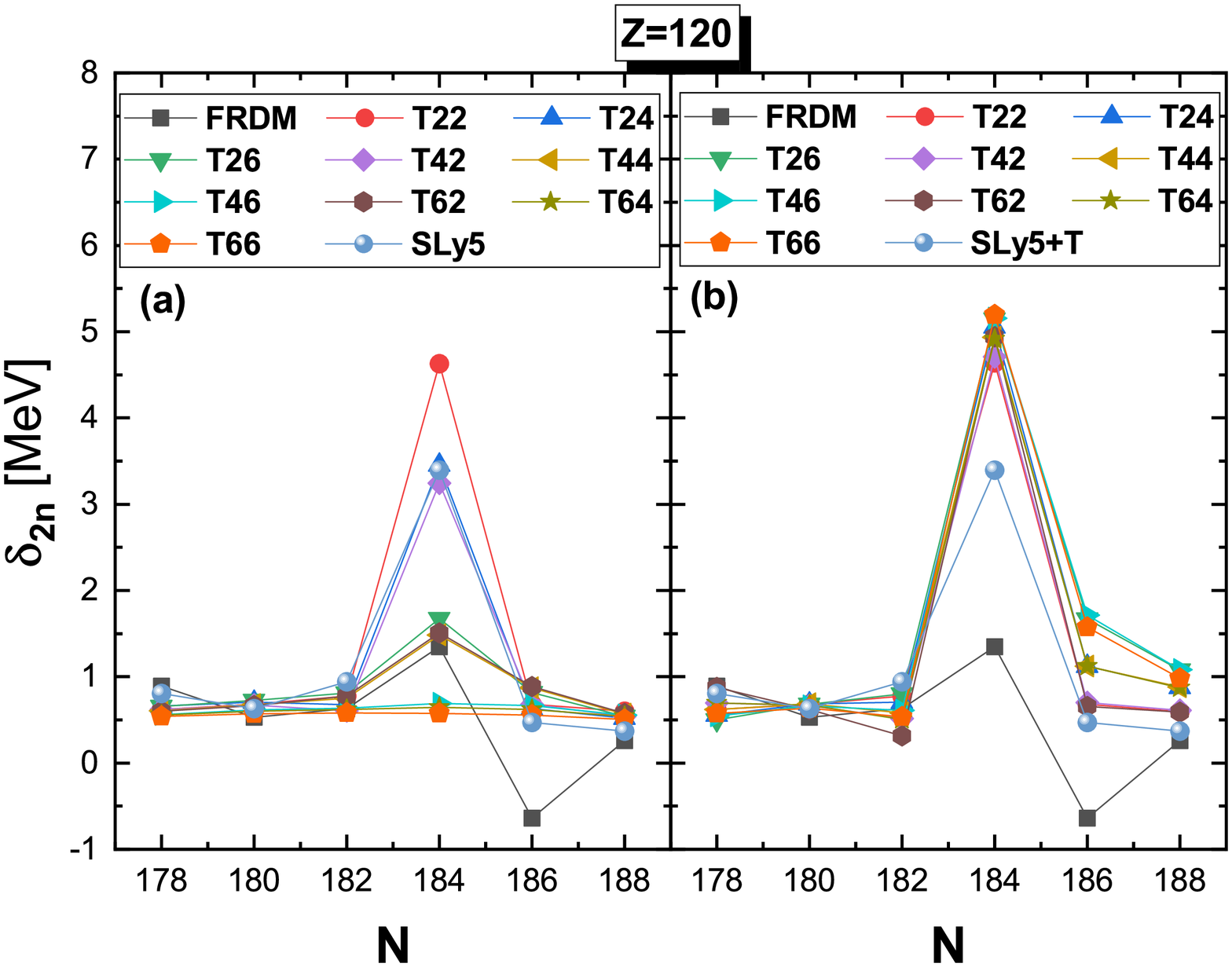}
	\end{minipage}\hfill
	\begin{minipage}{0.5\textwidth}
		\centering \includegraphics[scale=0.3]{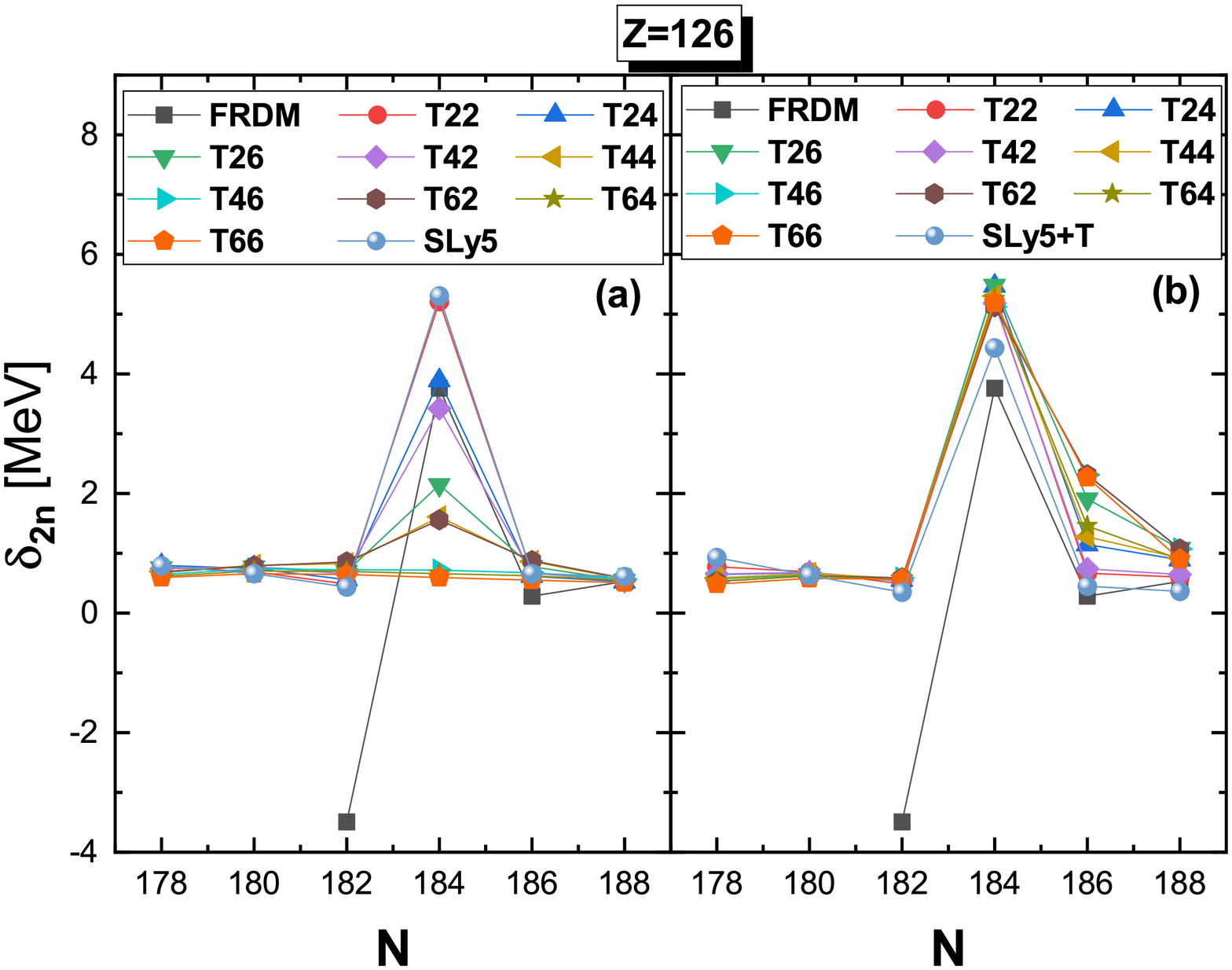}
	\end{minipage}\hfill
	\caption{The variations of two-neutron shell gap $\delta_{2n}(Z,N)$ across Z=114, Z=120 and Z=126 isotopes as a function of neutron number N in the cases with (panels (a)) and without (panels (b)) tensor force, using ten different Skyrme parametrizations. We compare our results with FRDM predictions\label{D2nFig}.}
\end{figure*}

The discovery and analysis of the shell gaps of
super-heavy nuclei is frequently based on the so-called two-neutron shell gap $\delta_{2n}(Z,N)$ that is defined as
\begin{eqnarray}
\delta_{2n}(Z,N)=S_{2n}(Z,N)-S_{2n}(Z,N+2)
\end{eqnarray}
This entity is considered as a more sensitive indicator to quantify and detect the shell closures than $S_{2n}(Z,N)$.  
In Fig. \ref{D2nFig}, we illustrate the variations of the calculated quantity  $\delta_{2n}(Z,N)$ with HF+BCS method in comparison with FRDM model. As can be seen in this Figure, there are peaks in $\delta_{2n}$ at neutron number N=184 for all selected parametrizations in both cases with  and without tensor force except for T46, T64 and T66 which exhibit no peaks at this number where the tensor force is absent. This can be explained by the same reason pointed out in the previous subsection for the quantity $S_{2n}$, i.e. the tensor force impact is more evident for large $C_0^J$.

It is also observed that the magnitude of the peaks is large and the results concord between each other when the tensor part is introduced. However, the inverse occurs with SLy5. Note that FRDM model has negative values in $\delta_{2n}(Z,N)$ at (Z=114, N=188), (Z=120 , N=186) and (Z=126, N=182) which is contradictory with the definition of this entity which should be positive. A part this, the neutron gap at N=184 is reproduced with this model. 

From a quantitative point of view, the tensor part leads to an increase in $\delta_{2n}(Z,N)$ that depends on the type of Skyrme sets (A part SLy5 and T22). Across the three super-heavy nuclei (Z=114, N=184), (Z=120 , N=184) and (Z=126, N=184), the increase is about 5.5MeV for T46, T64 and T66, 3.5MeV for T26, T44 and T62 and 2MeV for T24 and T42. However, for SLy5 the value of $\delta_{2n}(Z,N)$ is reduced by approximately 1.5MeV, while there is no change for T22. For all used parametrizations, it is clear that only the values of $\delta_{2n}(Z,N)$ at N=184 neutron shell gaps are influenced by introducing the tensor effect.

\subsection{Neutron pairing energy}
\begin{figure*}[tbh]
	\begin{minipage}{0.5\textwidth}
		\centering \includegraphics[scale=0.3]{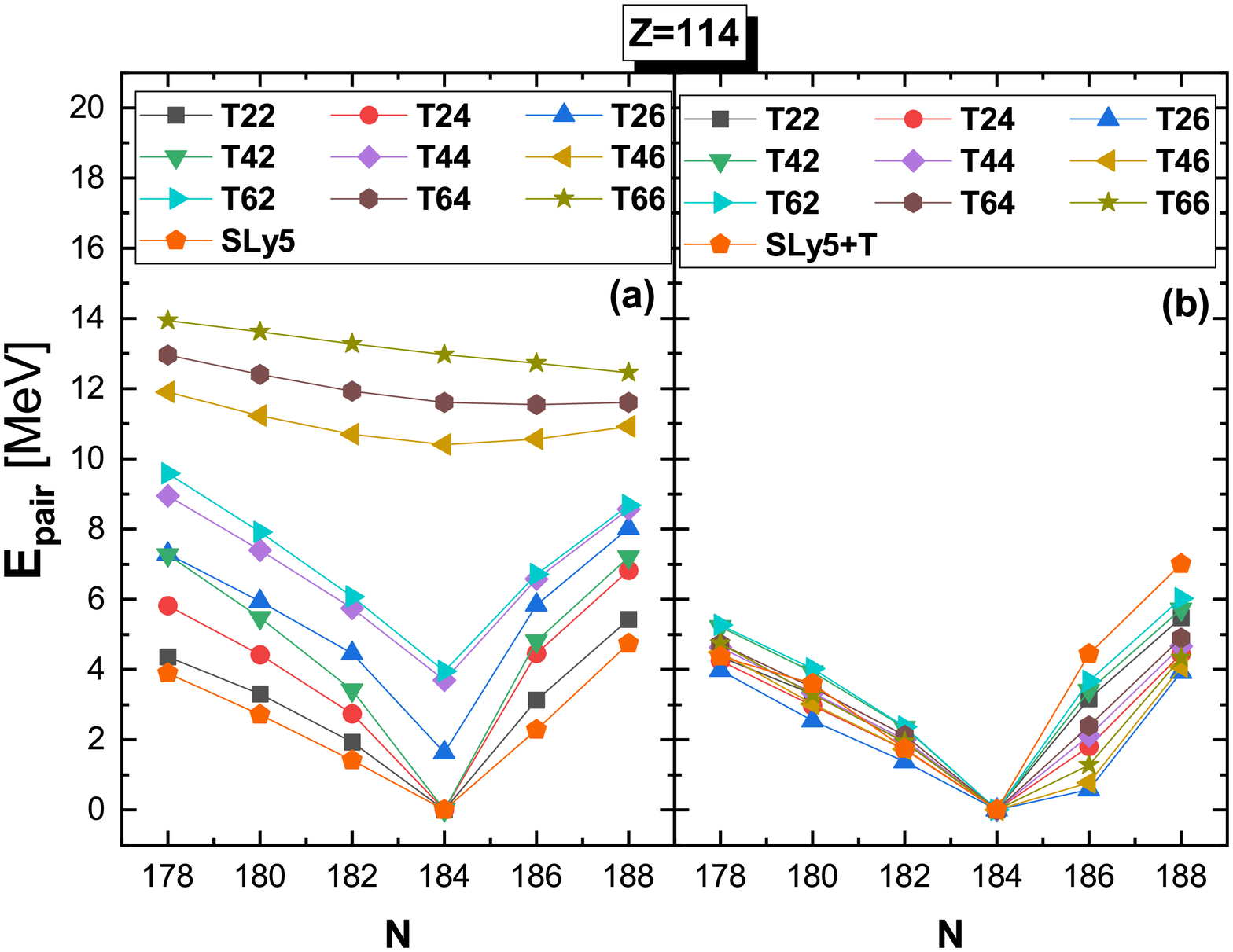}
	\end{minipage}
	\begin{minipage}{0.5\textwidth}
		\centering \includegraphics[scale=0.3]{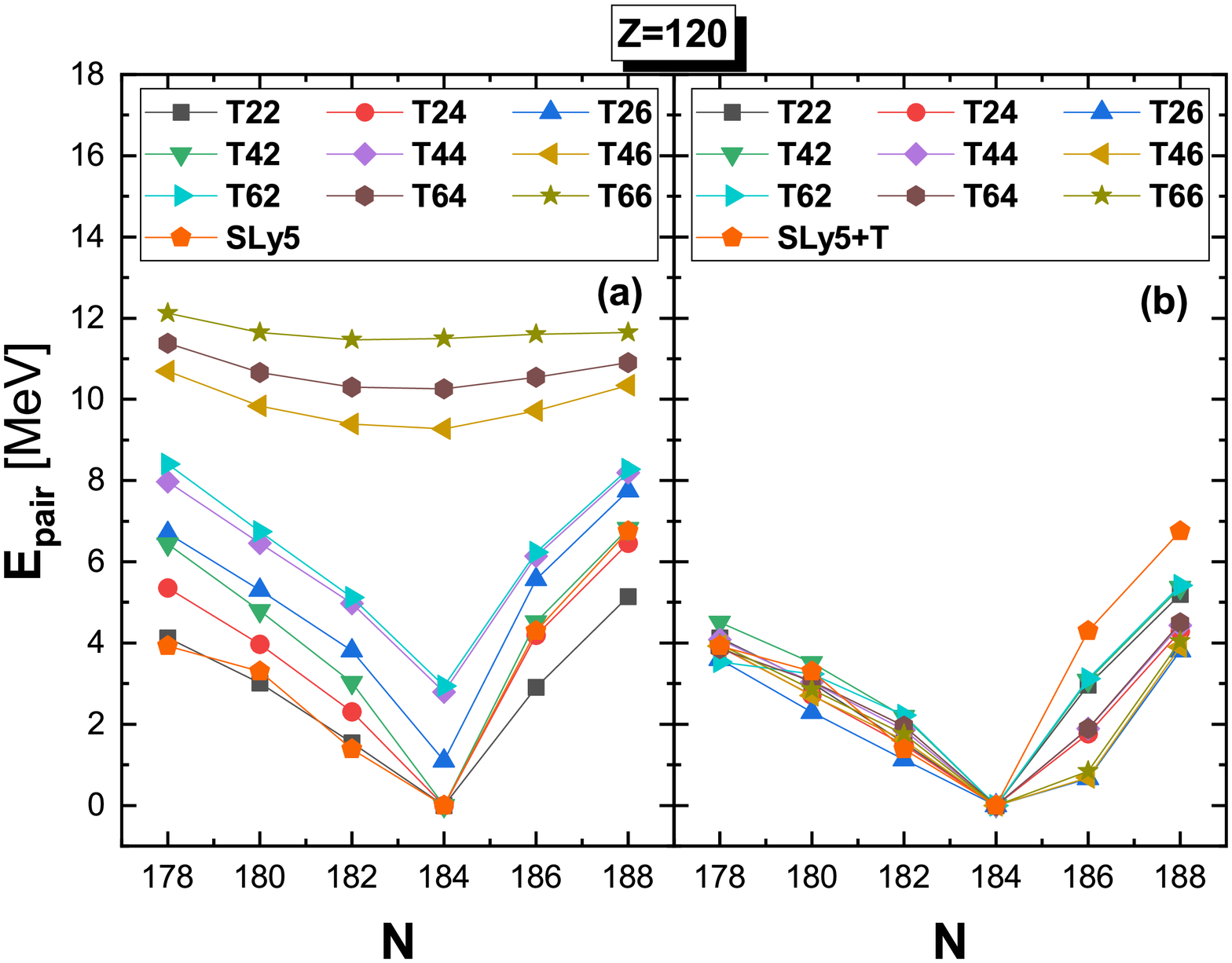}
	\end{minipage}\hfill
	\begin{minipage}{0.5\textwidth}
		\centering \includegraphics[scale=0.3]{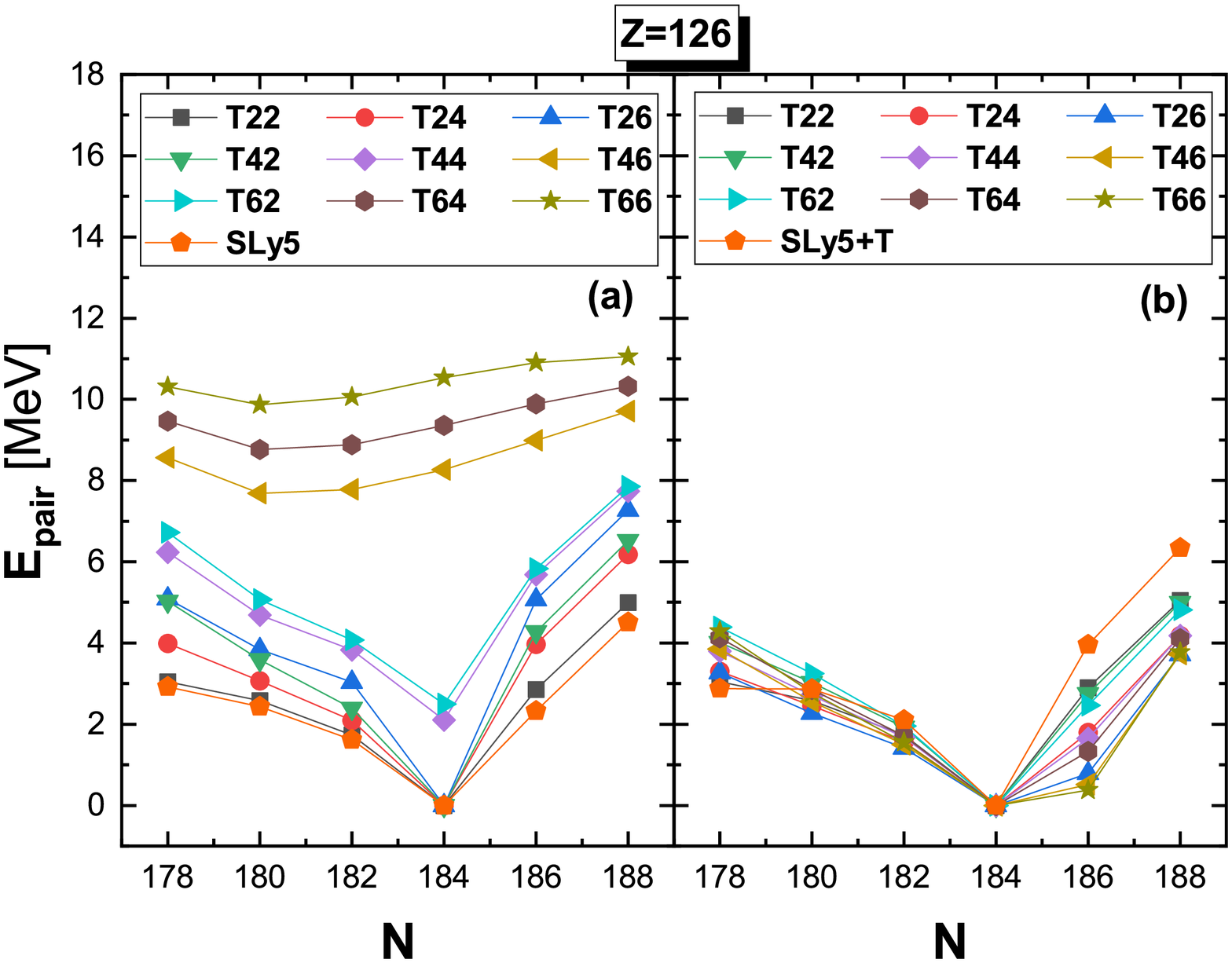}
	\end{minipage}\hfill
	\caption{Neutron pairing energy of Z=114, Z=120 and Z=126 elements calculated by HF+BCS approximation. The parameter sets SLy5, SLy5+T, T22, T24, T26, T42, T44, T46, T62, T64 and T66 are used. the panels (a) and (b) refer to calculated $E_{pair}$ quantity without and with the tensor force term, respectively. \label{EpairFig}}
\end{figure*}

In magic nuclei, the pairing effect is vanished. Thus, calculating the pairing energy of nuclei allows to detect their magicity. In Fig. \ref{EpairFig}, we display the neutron pairing energy $E_{pair}$ for the even-even Z=114, Z=120 and Z=126 isotopes calculated by HF+BCS using ten different Skyrme parameters SLy5, T22, T24, T26, T42, T44, T46, T62, T64 and T66 as well as the predictions of FRDM approach. The panels (a) show the obtained results without tensor force and the panels (b) illustrate those with tensor force. In both cases, the values of neutron pairing energy repeat the predictions as achieved by the the quantities $S_{2n}$ and $\delta_{2n}$ and show nice agreement between results obtained from all Skyrme parametrizations as seen in panels (b) of Fig. \ref{EpairFig}.\\
If we study the $E_{pair}$ curves carefully, it is clear that this quantity vanishes at the neutron magic number N=184 in the case of the force parametrizations T22, T24, T42 and SLy5, with a lower coupling constant $C_0^J$, for all three elements Z=114, Z=120 and Z=126 in both cases with and without tensor force. For  T26, T44, T46, T62, T64 and T66, only in the case where the tensor force is included, the $E_{pair}$ vanishes. In the case without tensor force, the sets with medium values of coupling constant $C_0^J$, i.e. T26, T44 and T62, have minimum values for $E_{pair}$ at N=184, but do not reach zero. However, the $E_{pair}$ based on the sets T46, T64 and T66 with higher $C_0^J$ varies linearly, showing no shell closure at N=184. Here, also the discussions of pairing energy allow us to conclude that the tensor force effect strongly depends on the values of $C_0^J$.\\
\subsection{Rn and Rp}
\begin{figure*}[tbp]
	\begin{minipage}{0.5\textwidth}
		\centering \includegraphics[scale=0.3]{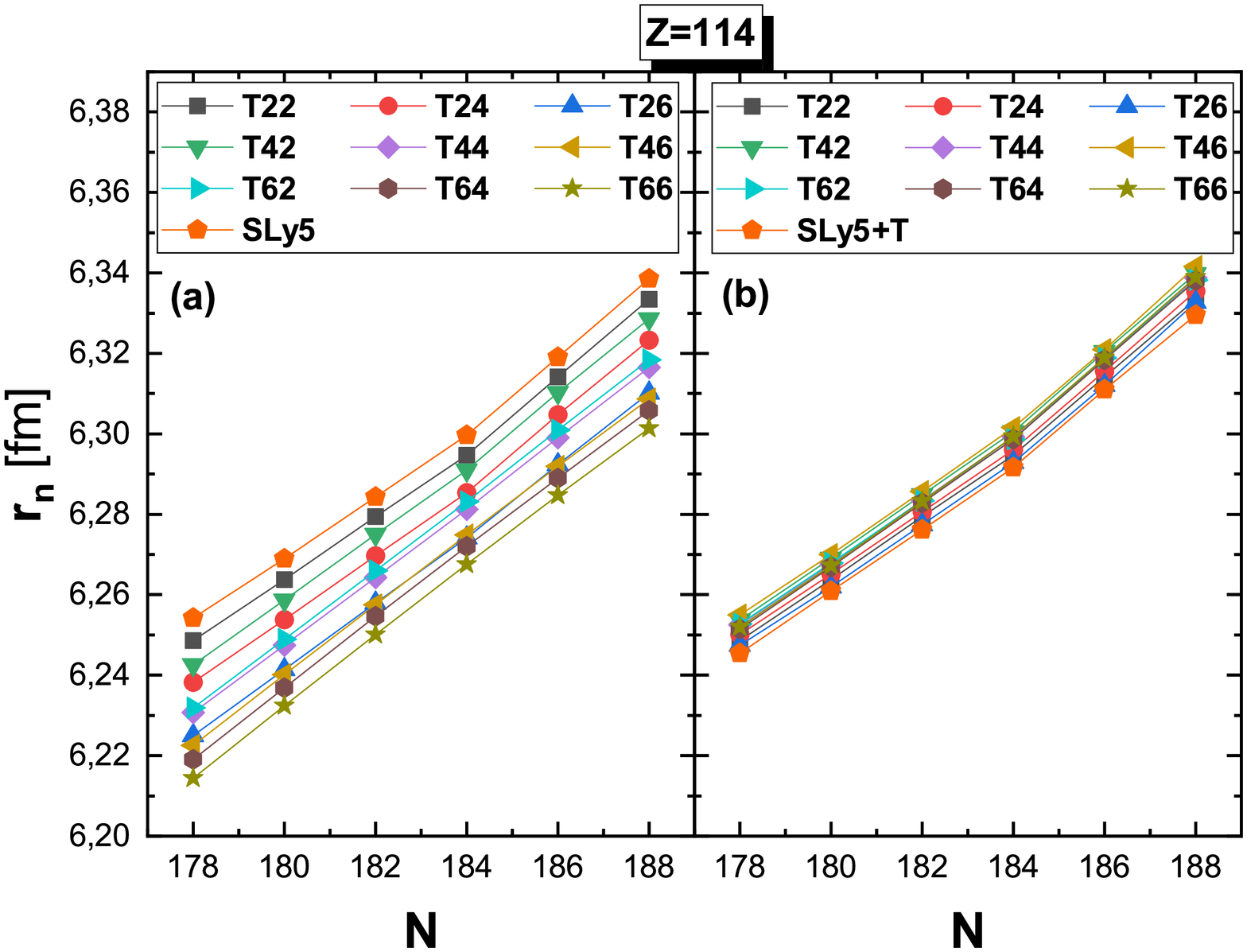}
	\end{minipage}
	\begin{minipage}{0.5\textwidth}
		\centering \includegraphics[scale=0.3]{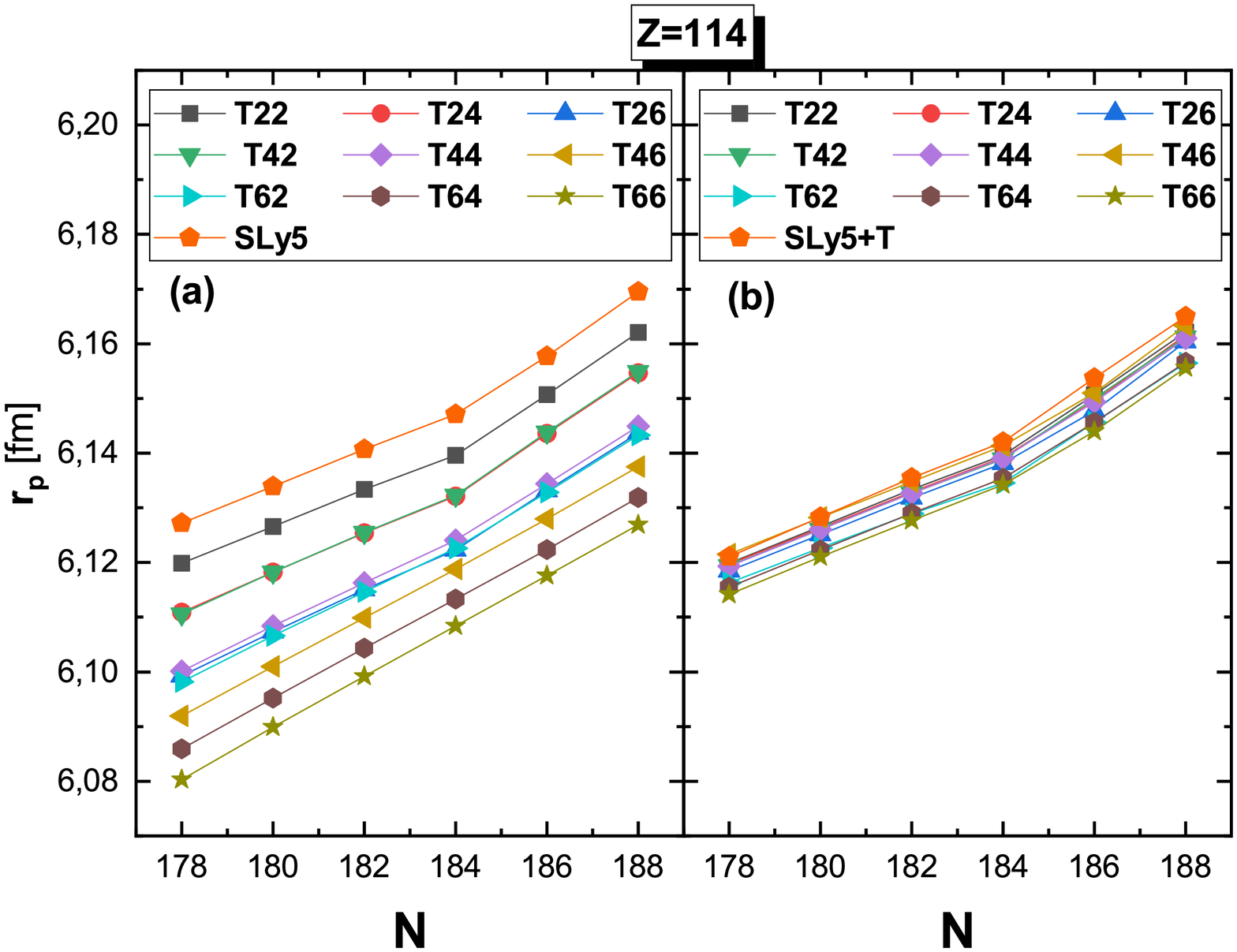}
	\end{minipage}\hfill
	\begin{minipage}{0.5\textwidth}
		\centering \includegraphics[scale=0.3]{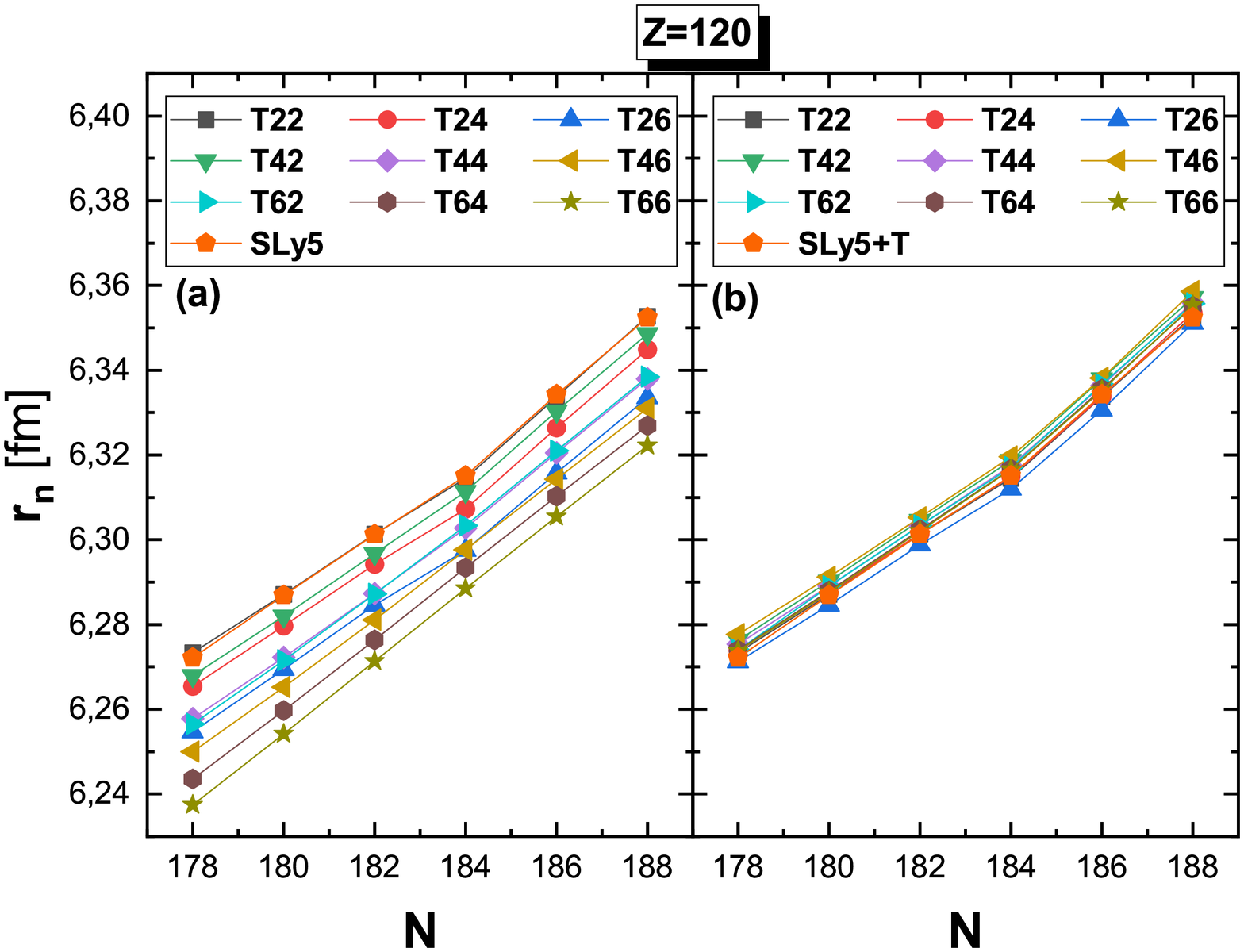}
	\end{minipage}
	\begin{minipage}{0.5\textwidth}
		\centering \includegraphics[scale=0.3]{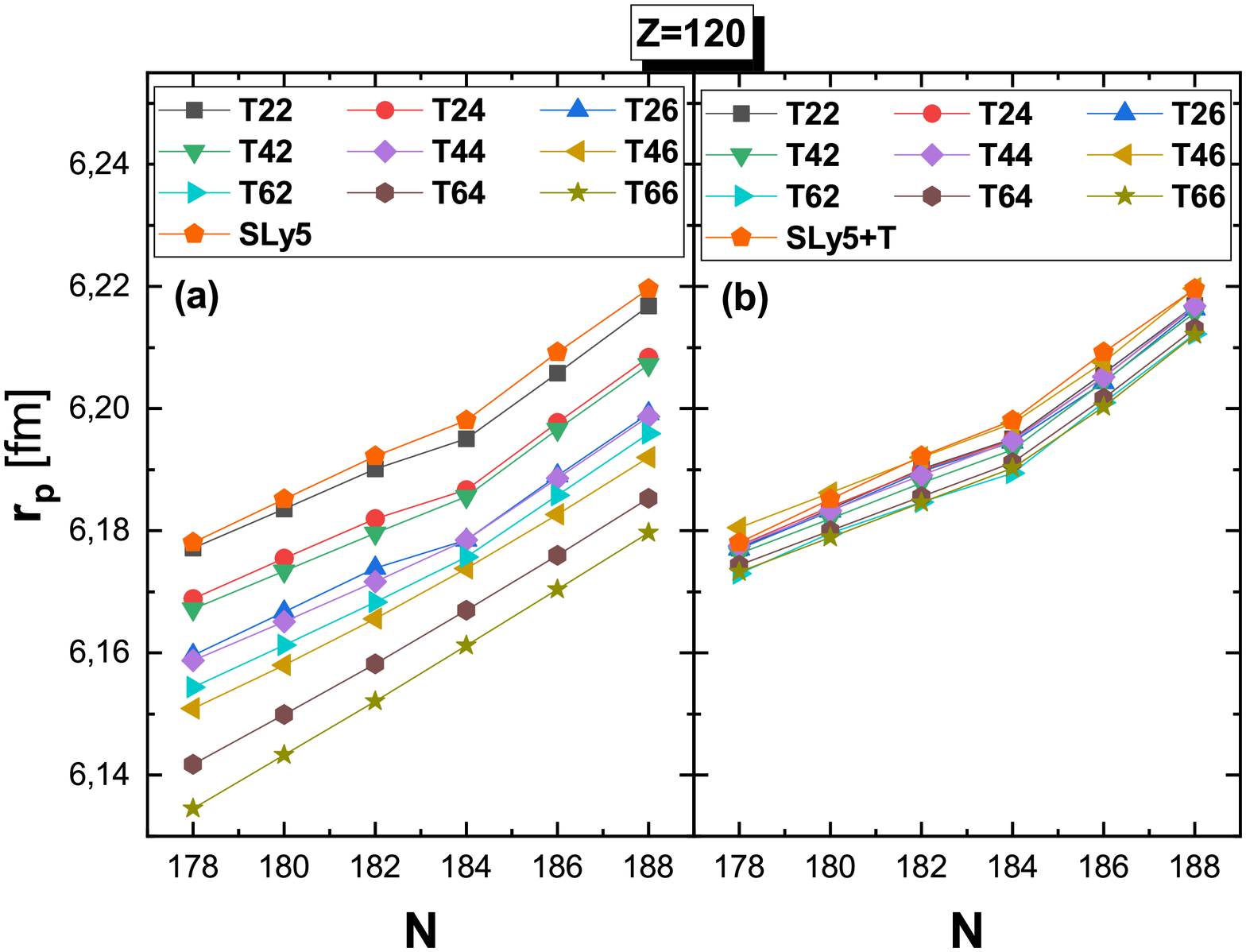}
	\end{minipage}\hfill
	\begin{minipage}{0.5\textwidth}
		\centering \includegraphics[scale=0.3]{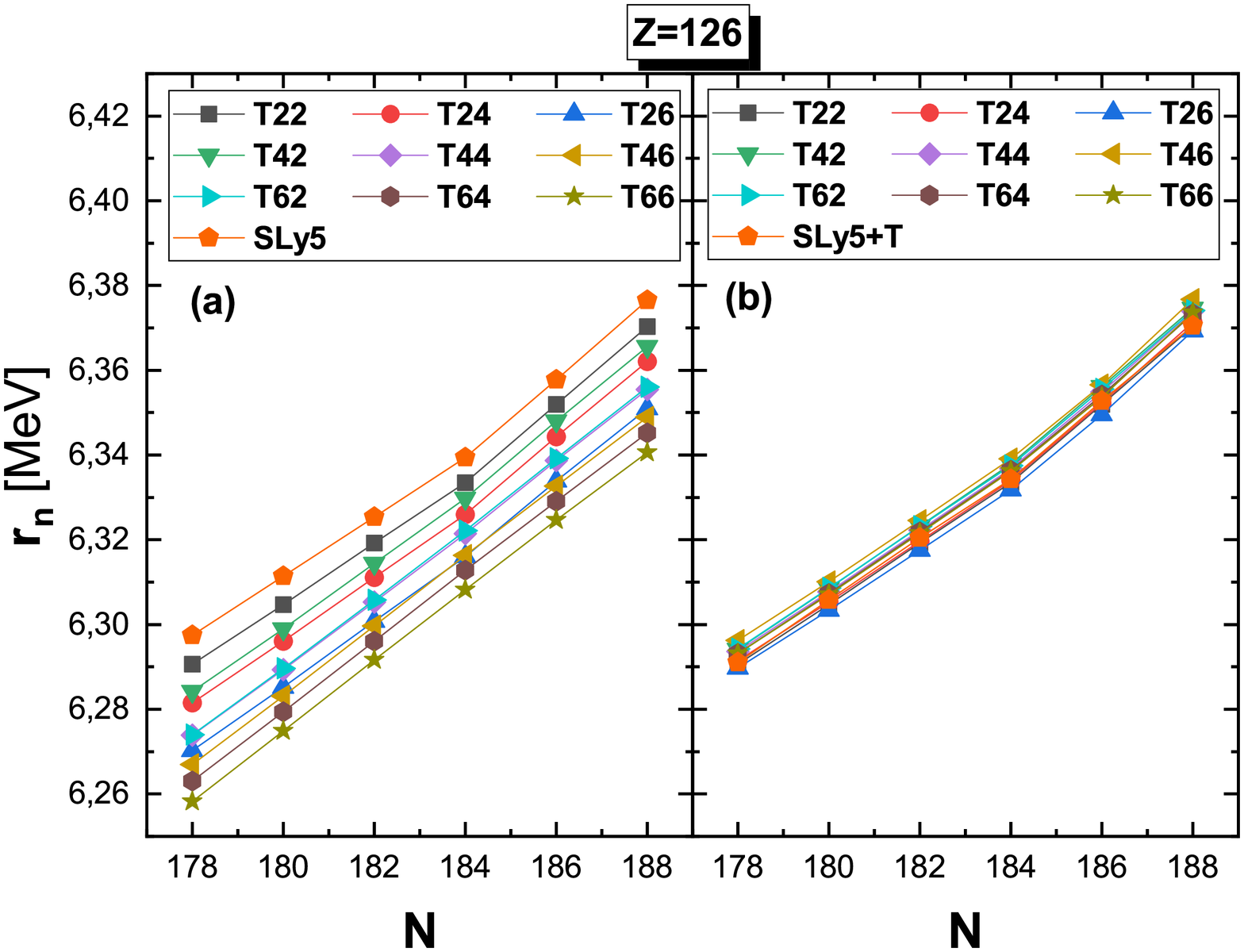}
	\end{minipage}
	\begin{minipage}{0.5\textwidth}
		\centering \includegraphics[scale=0.3]{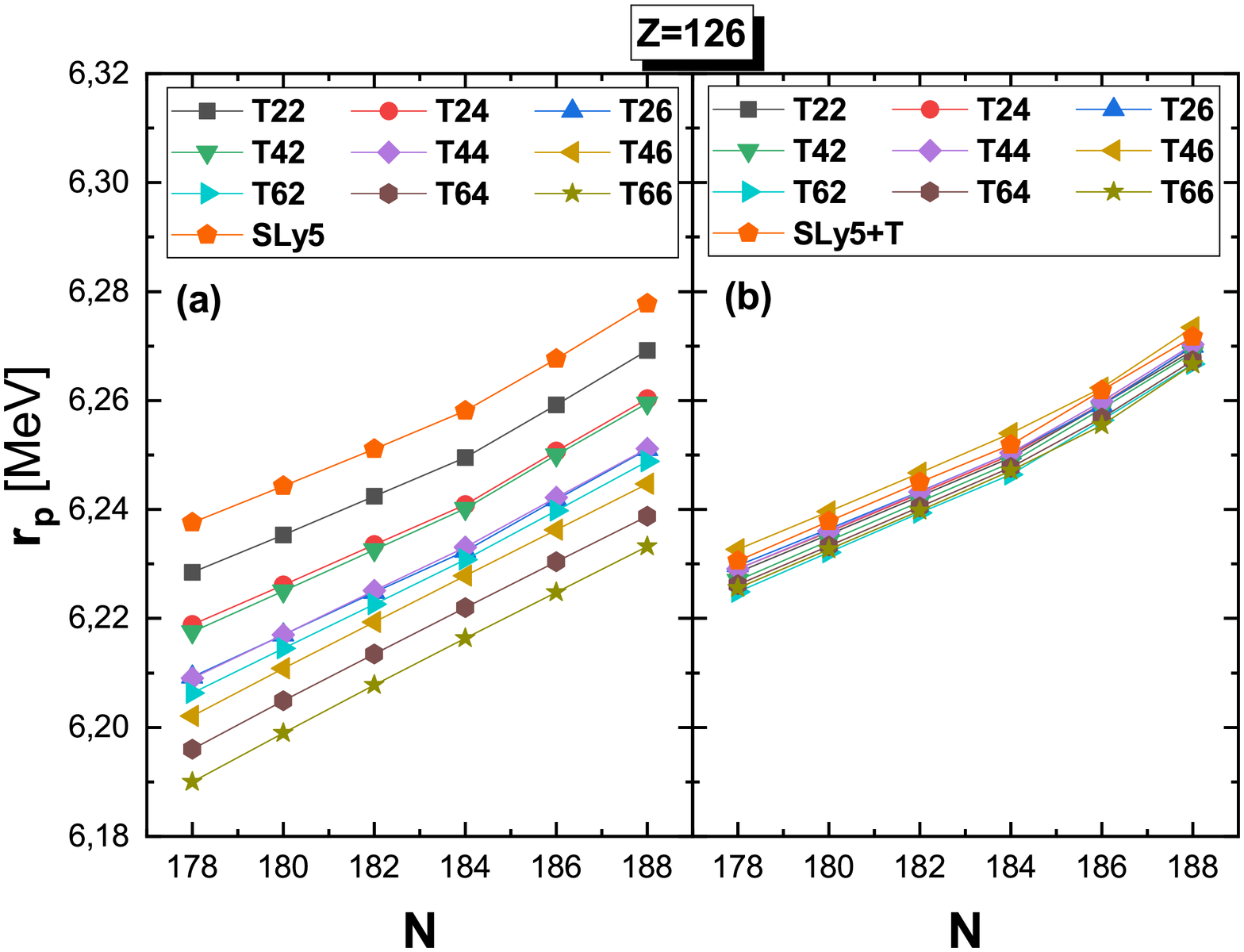}
	\end{minipage}\hfill
	\caption{Root mean-square radii for neutrons (left curves) and protons (right curves) obtained by means of HF+BCS model for the even-even isotopes with Z=114, 120 and 126. In the panels (a) and (b) we show the results obtained without and with the tensor interaction, respectively.\label{RmsFig}}
\end{figure*}

The mean-square (ms) radii for  neutron (q=n) and proton (q=p) distributions are defined by
\begin{eqnarray}
r^2_q=\frac{1}{N_q}\int d^3r r^2 \rho_q(\mathbf{r})
\end{eqnarray}
and the root mean-square (rms) radii are given by
\begin{eqnarray}
r_q=\sqrt{r^2_q}
\end{eqnarray}
It is interesting to recall that rms radii may be used as an observable to detect shell closures. They present a kink around the magic numbers. 
The calculated  neutron $r_n$ (left panels) and proton $r_p$ (right panels) rms radii, in the HF+BCS approximation without tensor term (panels (a)) and with it (panels (b)), are shown in Fig. \ref{RmsFig}. For the three considered  elements, a dip in both neutron and proton radii at N=184 is seen for all used Skyrme parameters with the presence of the tensor force. When this force is not taken into account, only the SLy5, T22, T24, T26, T42, T44 and T62 sets exhibit a slight decrease in proton and neutron radii at N=184. For the remaining Skyrme sets, i.e. T46, T64 and T66, the behavior of these quantities presents no change, increasing as a finite function.\\
Another observation can be made from Fig. \ref{RmsFig}. In the case without tensor force, the selected sets present a large discrepancy in the radii values. This discrepancy varies between 0.04fm for neutrons and 0.05fm for protons. However, it is almost hidden when the tensor part is taken into account. 

Again, calculations show that the tensor component generally plays a crucial role in enhancing the prediction of the nuclei structure in the ground state. In particular for the sets with higher coupling constant $C_0^J$.
\subsection{Single particle energies}
\begin{figure*}[tbh]
	\begin{minipage}{0.5\textwidth}
		\centering \includegraphics[scale=0.3]{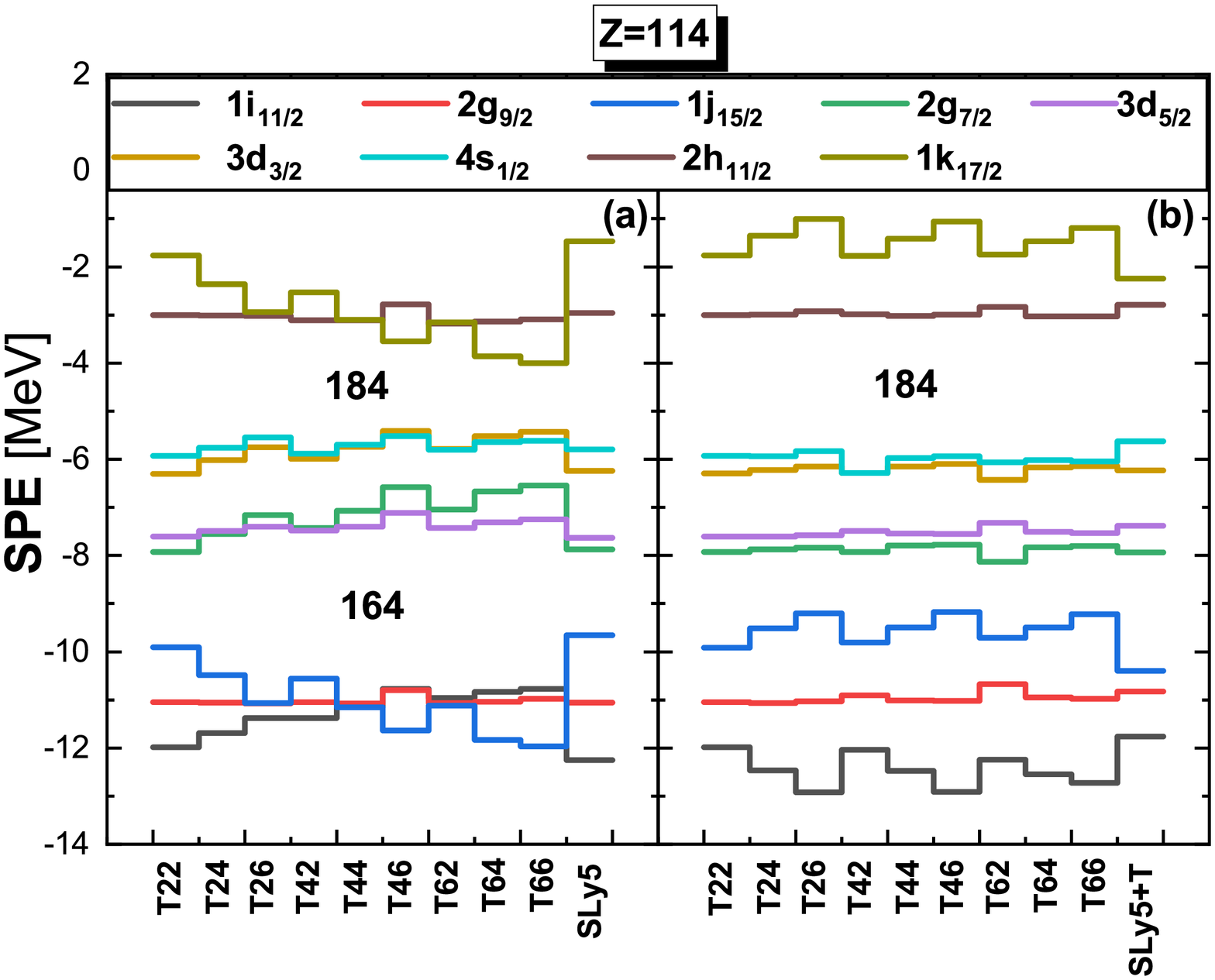}
	\end{minipage}
	\begin{minipage}{0.5\textwidth}
		\centering \includegraphics[scale=0.3]{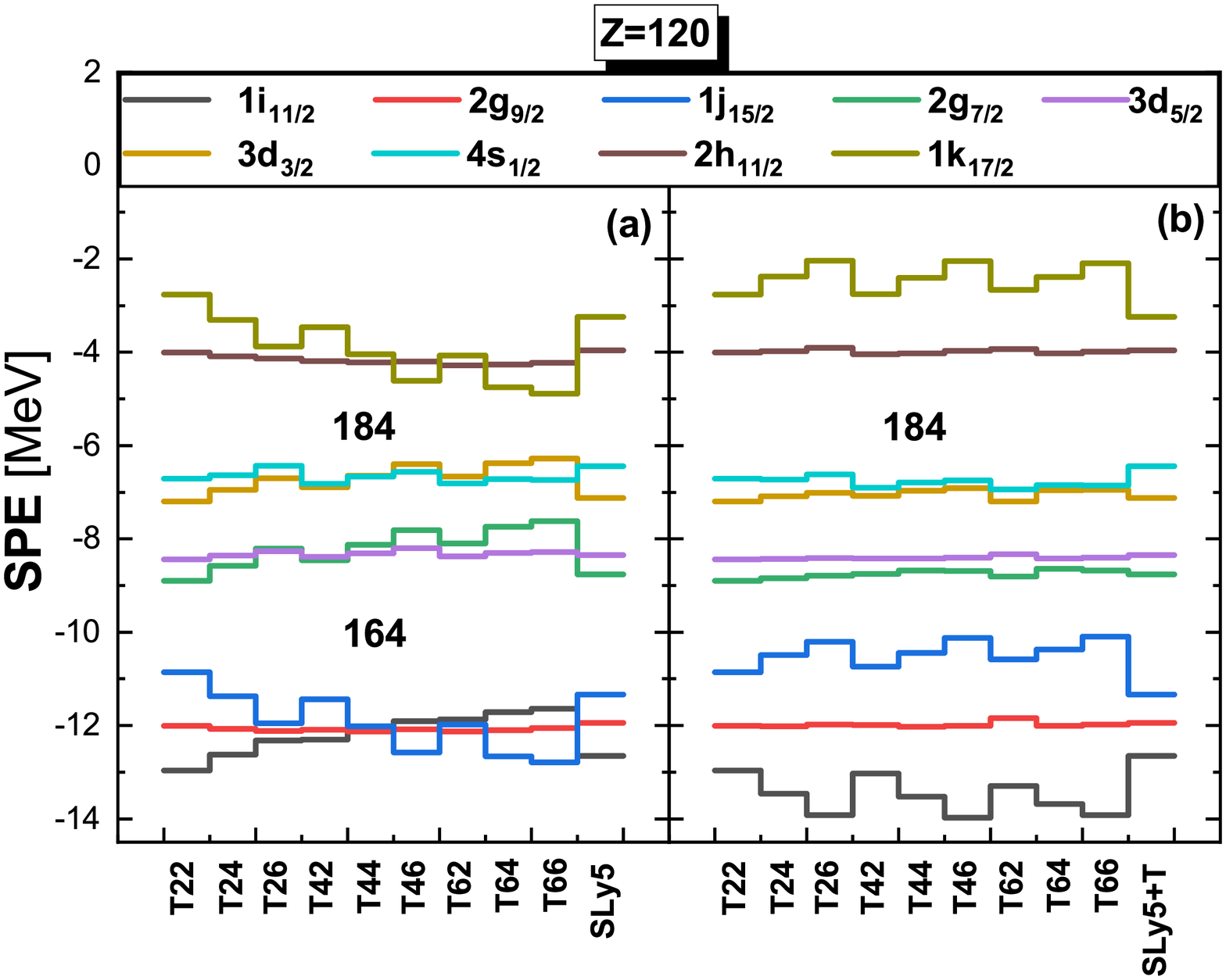}
	\end{minipage}\hfill
	\begin{minipage}{0.5\textwidth}
		\centering \includegraphics[scale=0.3]{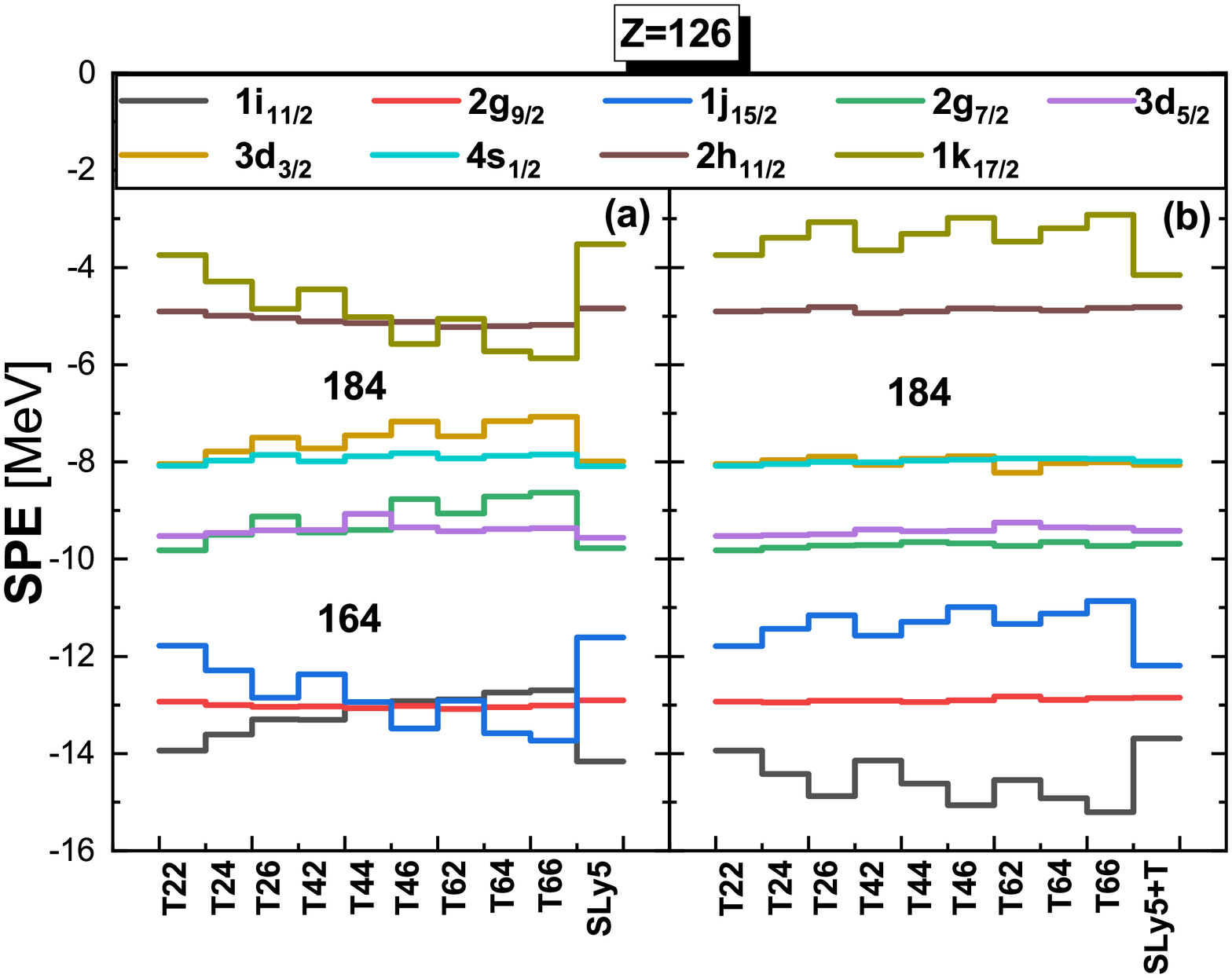}
	\end{minipage}\hfill
	\caption{Single-particle spectrum of the neutrons in $^{298}114$, $^{304}120$ and $^{310}126$ calculated with the HF+BCS approximation by means of several Skyrme functionals as indicated in x-axis, in the cases without (panels (a)) and with (panels (b)) tensor term.\label{SPEFig}}
\end{figure*}
In order to show more clearly the impact of the tensor force on the N=184 shell closure we plot in Fig. \ref{SPEFig} the sensitive observable to this force: single particle energies (SPE). The possible shell closure at N=184 is located between $4s_{1/2}$ and $2h_{1/2}$ levels. 
From the panels (a) of Fig. \ref{SPEFig}, a strong N=184 gap is observed only for T22 and SLy5 (with lower $C_0^J$ compared to other functionals as given in Table \ref*{table}), when there is no contribution of the tensor term, in the three isotopic chains with Z=114, 120 and 126. But it is to be noticed that all other functionals under consideration overestimate the gap at N=164 and show at least a subshell closure at N=184 in all studied nuclei .

We now turn to study the spectrum of Z=114, 120 and 126 in the case where the contribution of the tensor force is taken into account. From the panels (b) of Fig. \ref{SPEFig}, the all Skyrme forces show a significant shell closure at N=184. The observed gap at N=164 in the case without tensor force is here significantly reduced. But only for SLy5, this gap is large enough to be treated as a major shell closure. Even so, the spectrum of N=184 in both cases is basically unchanged. Therefore, the gap at N=184 with SLy5 is somewhat influenced by the tensor force. This is due to the fact that this force is included into the existing SLy5 set compared to TIJ functionals which are built with a complete fit of all parameters.

It is worth mentioning that the single-particle spectra of the stats $2g_{9/2} $, $3d_{5/2} $, $4s_{1/2} $ and $2h_{11/2}$, of all functionals, have almost the same values as well as other remaining neutrons states of T22 and SLy5 in both cases with and without tensor force. Furthermore, The $2g_{9/2} $, $3d_{5/2} $, $4s_{1/2} $ and $2h_{11/2}$ states look independent to Skyrme forces.

It is also observed from Fig. \ref{SPEFig} that the order of the neutron single-particle energies is not the same comparing the cases with and without tensor force for all Skyrme interactions with exception of T22, T24 and SLy5. In some cases, the forces T26, T42 and T44 reproduce the same order of single-particle spectra with and without tensor term. we quote for example, the case of $1k_{17/2} $, $4s_{1/2} $ and $1j_{17/2} $ states.

To sum up, the tensor term again leads to a closure shell at N = 184, which is of a similar magnitude for most forces. Only for T22 and SLy5 functionals with a lower coupling constant $C_0^J$, the tensor term produces small changes in the spin-orbit splitting around this gap. However, with an increase of $C_0^J$, its effect  became large enough to modify the shell gap N=184 for other remaining Skyrme forces. This is obviously induced by the different values of the coupling constant $C_0^J$.

\subsection{$\alpha$-decay properties}  
\begin{figure*}[tbh]
	\begin{minipage}{0.5\textwidth}
		\centering \includegraphics[scale=0.3]{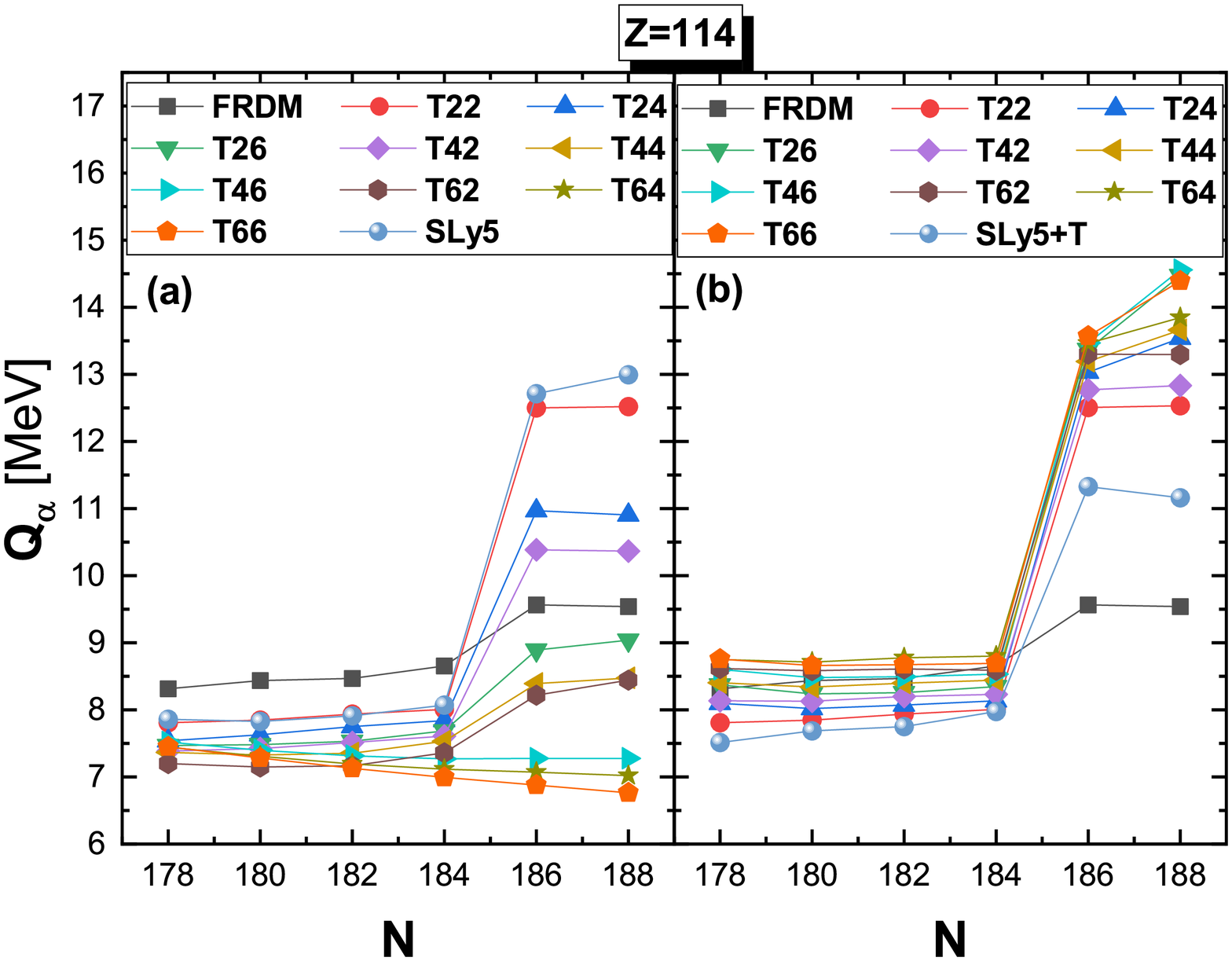}
	\end{minipage}
	\begin{minipage}{0.5\textwidth}
		\centering \includegraphics[scale=0.3]{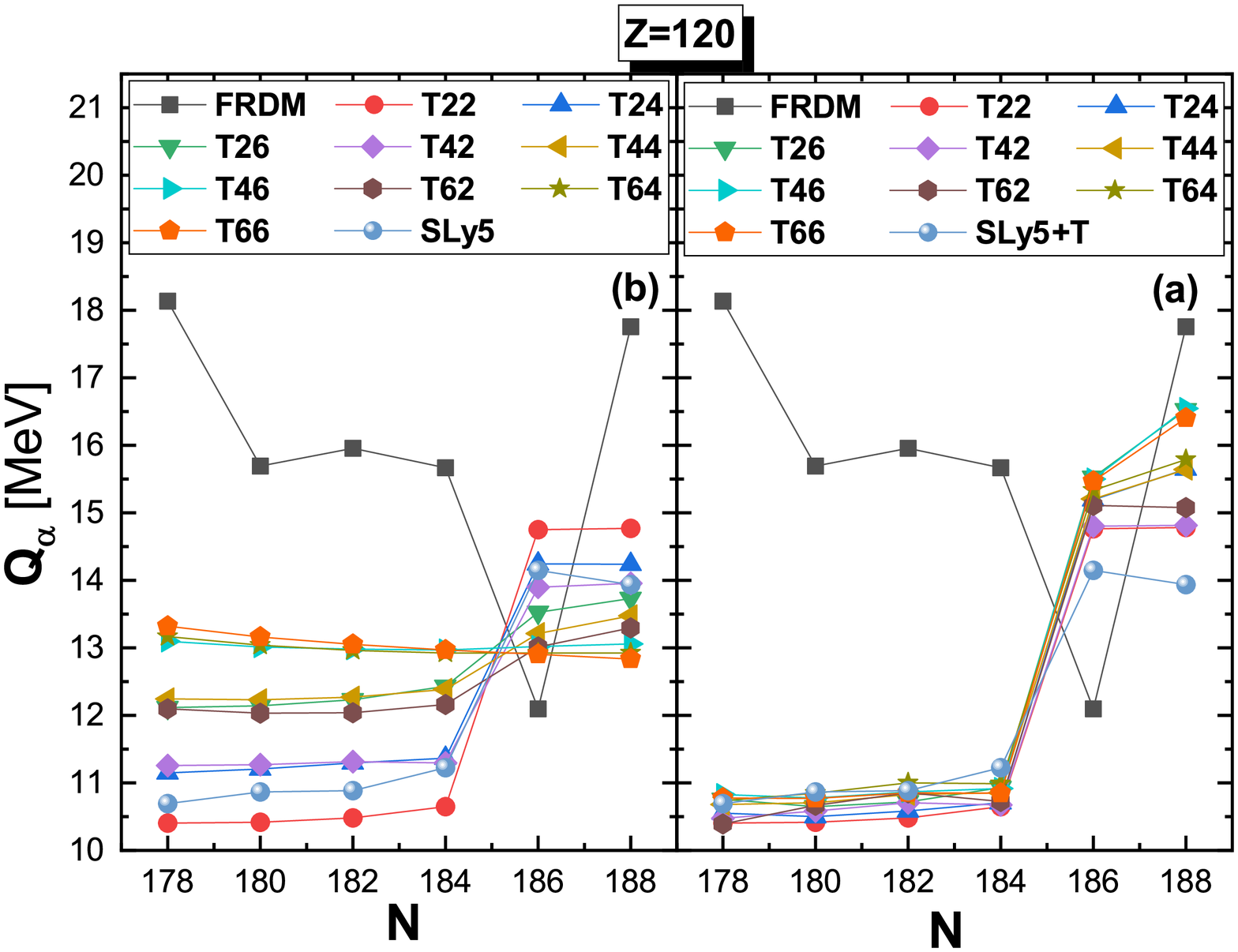}
	\end{minipage}\hfill
	\begin{minipage}{0.5\textwidth}
		\centering \includegraphics[scale=0.3]{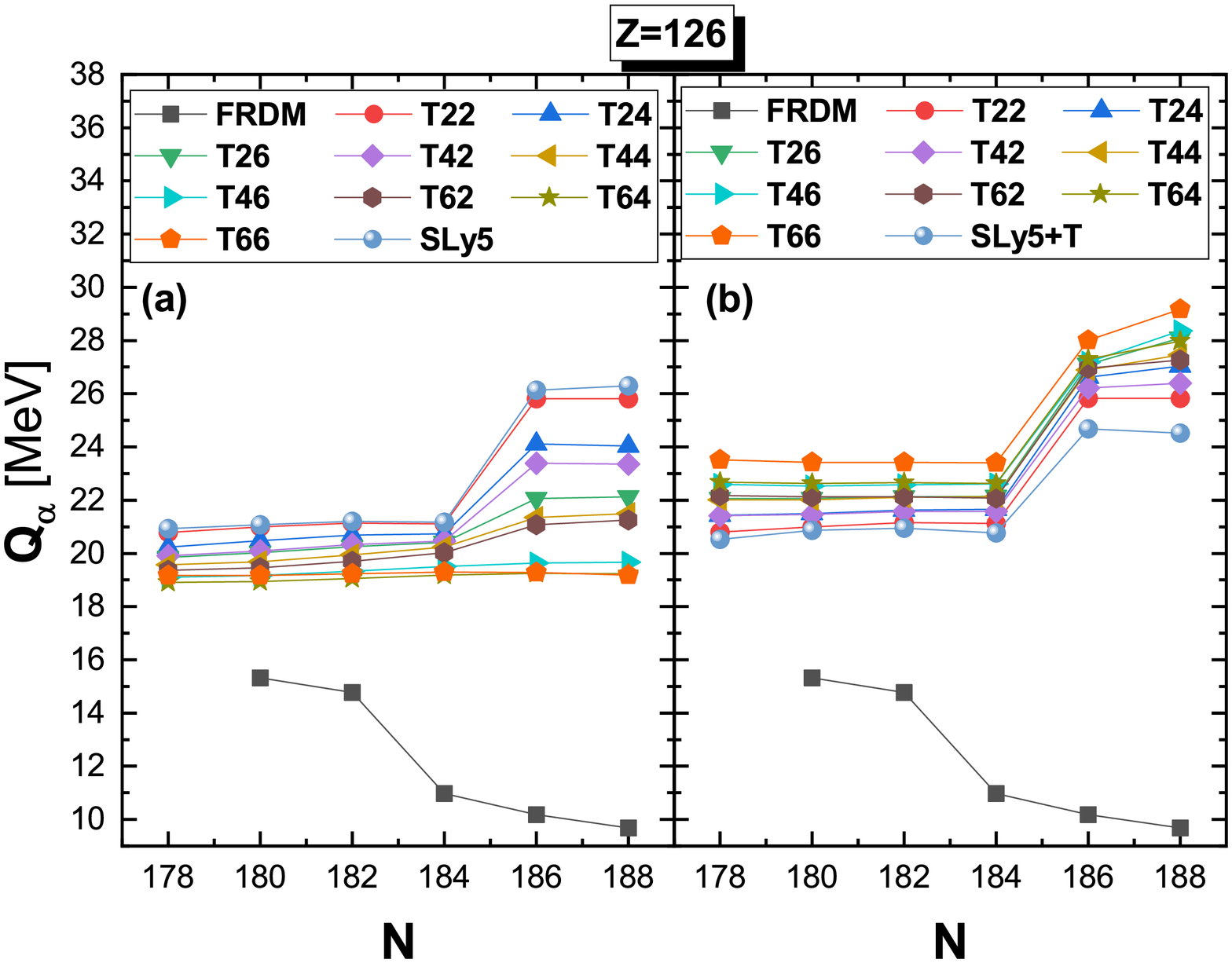}
	\end{minipage}\hfill
	\caption{The calculated $Q_{\alpha}$ values for even-even super-heavy nuclei with Z=114, Z=120 and Z=126. In the panels (a) and (b) we show the results obtained without and with the tensor interaction, respectively. Our results are compared with FRDM predictions.\label{Q_aFig}}
\end{figure*}
In super-heavy nuclei, there is a competition between two decay modes: the spontaneous fission and $\alpha$-decay. These decay modes allow to determine the possibility of existence of SHN as well as the degree of their stability. In experiments, the observation of SHF is not possible unless the half-life of spontaneous fission  is longer than the $\alpha$-decay half-life.  Anyway, that observation is only possible if half-lives are greater than $\tau=10\mu s$.

As it is known, The $\alpha$-decay half-life depends on the $\alpha$-decay energy $Q_{\alpha}$ which is calculated from the binding energies BE(N, Z), BE (N-2, Z-2) and BE (2,2)=28.296 MeV \cite{Wang2012} of the parent, daughter and $\alpha$ particle, respectively, according to the following relation
\begin{eqnarray}
Q_{\alpha}=BE(N,Z)-BE(N-2,Z-2)-BE(2,2)
\end{eqnarray}
Note that  the $\alpha$-decay is possible only if $Q_{\alpha}$ has positive value.\\
Using $Q_{\alpha}$ values, we computed the $\alpha$-decay half-lives with a new phenomenological formula \cite{moumen}
\begin{eqnarray}
log_{10}\tau_{\alpha} & = & \frac{0.27Z-0.54}{\sqrt{Q_{\alpha}}}-\frac{e(Z-1)}{Q_{\alpha}^f}+(Z-1)\\ \nonumber &  &
\Big[0.22A^{1/6}-a-bA+cZ^d+gA^{1/2}\Big]-48.72
\end{eqnarray}
with the parameters $a=3.064$, $b=0.008$, $c=7.872 10^{-6}$, $d=2.31$, $e=-3.54$, $f=0.451$ and $g=0.218$ \cite{moumen}.

\begin{figure*}[tbh]
	\begin{minipage}{0.5\textwidth}
		\centering \includegraphics[scale=0.3]{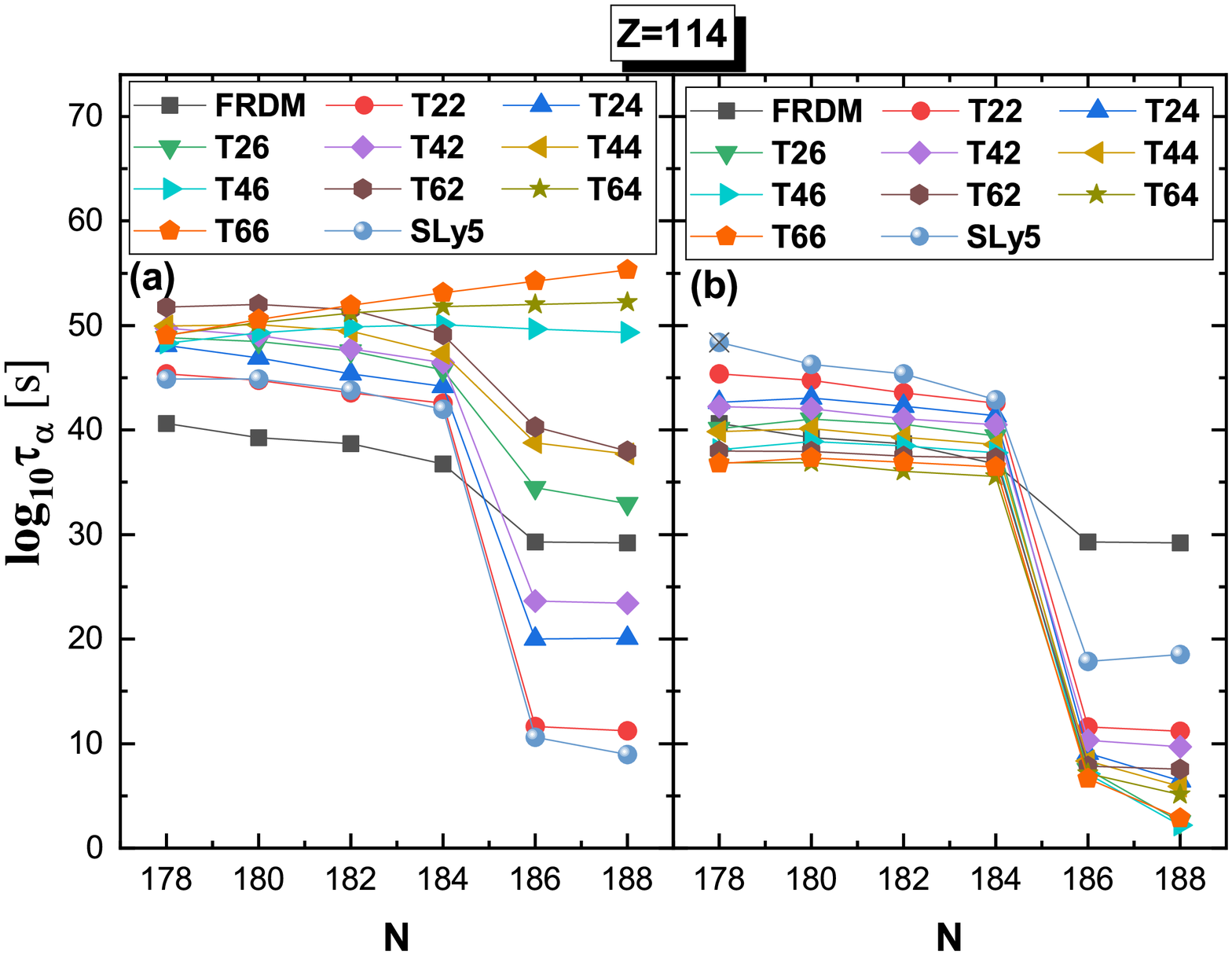}
	\end{minipage}
	\begin{minipage}{0.5\textwidth}
		\centering \includegraphics[scale=0.3]{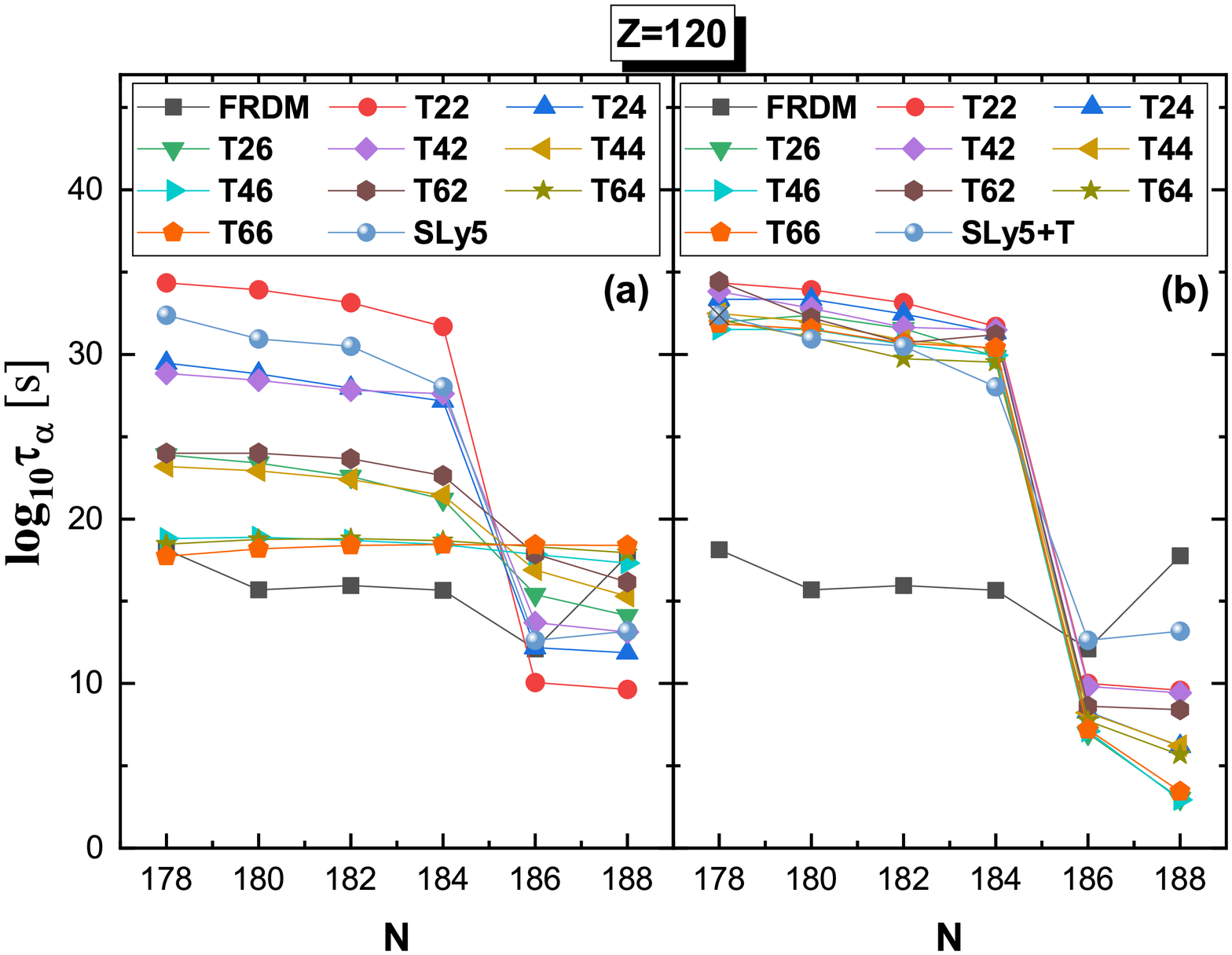}
	\end{minipage}\hfill
	\begin{minipage}{0.5\textwidth}
		\centering \includegraphics[scale=0.3]{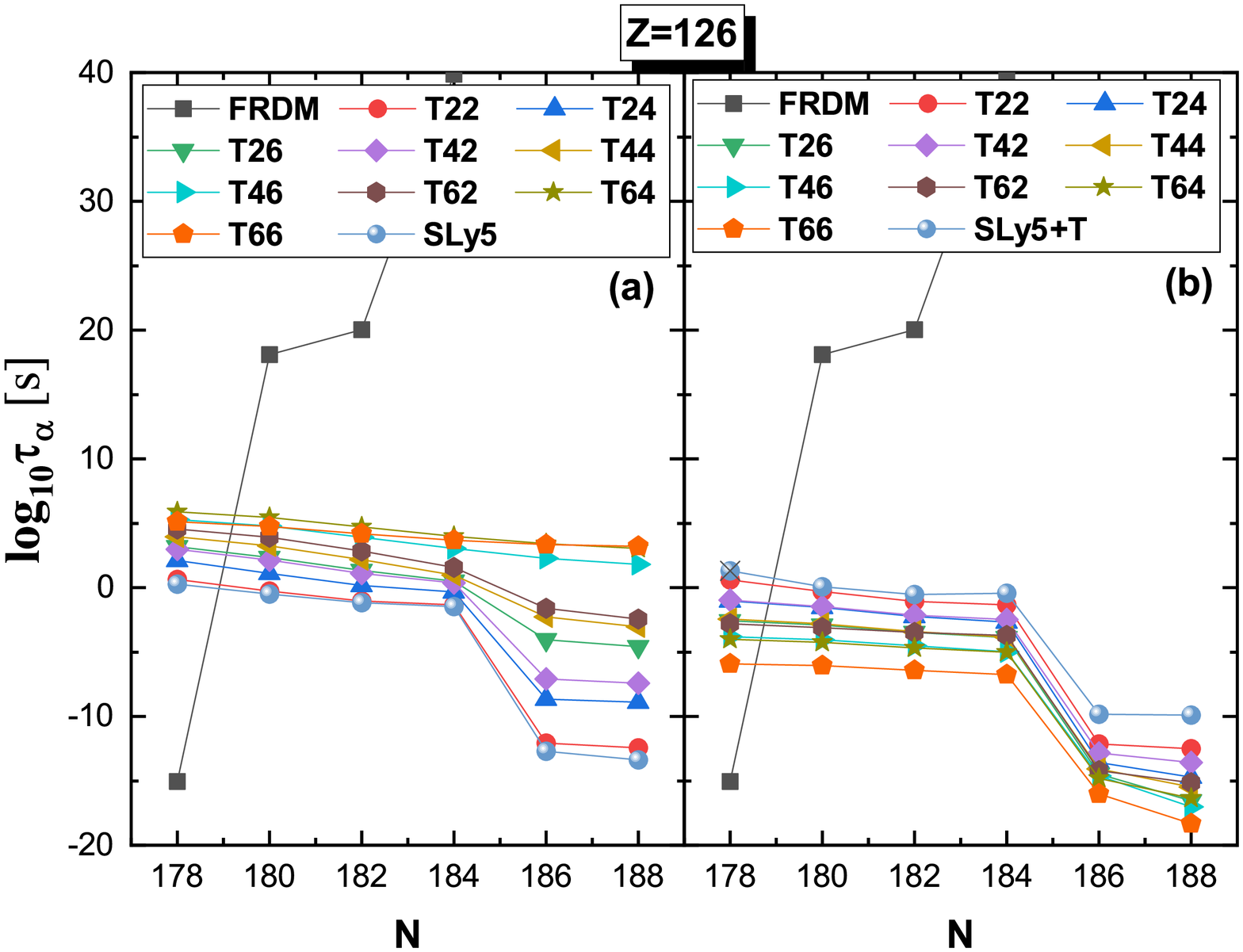}
	\end{minipage}\hfill
	\caption{The calculated half-lives for $\alpha$-decays of even-even super-heavy nuclei with Z=114, Z=120 and Z=126. In the panels (a) and (b) we show the results obtained without and with the tensor interaction, respectively. Our results are compared with FRDM predictions.\label{T_aFig}}
\end{figure*}

The calculated quantities $Q_{\alpha}$ and $\alpha$-decay half-lives, using the SLy5, T22, T24, T26, T42, T44, T46, T62, T64 and T66 parametrizations with (panels (b)) and without tensor force (panels (a)), for the isotopic chains of Z=114, 120 and 126, are shown graphically in Fig. \ref{Q_aFig} and Fig. \ref{T_aFig}, respectively. Comparison is done with FRDM predictions.\\
In Fig. \ref{Q_aFig} and Fig. \ref{T_aFig}, there is a sharp peak at N=184 in $Q_{\alpha}$-energy and a local decrease in $log_{10}\tau _{\alpha}$ for all Skyrme sets when the tensor interaction is introduced, suggesting a stable system. Whereas the trend is different when this interaction is missed. As shown in panels (a), the N=184 shell closure is significant only with the SLy5, T22, T24 and T42 sets and less significant with T26, T44 and T62. For T46, T64 and T66, the $Q_{\alpha}$-energy and $\alpha$-decay half-lives present an almost constant variation as functions of the number of neutrons. According to the above discussion, it is easy to see that the tensor force effect on the shell closure evolution at N=184 is also sensitive to the coupling constant $C_0^J$.
Our numerical results are compared with the theoretical predictions of the FRDM model. From Fig. \ref{Q_aFig}, we can notice that the $Q_{\alpha}$-energy values from FRDM are closer to those obtained using Skyrme functionals for the isotopic series of Z=114. But, for Z=120 and Z=126 isotopes, the situation gets opposite and FRDM shows, at N=184, an decreasing trend with increasing neutron numbers rather than an increase in the case of Skyrme interactions. Similarly, curves of $log_{10}\tau _{\alpha}$ in Fig. \ref{T_aFig} exhibit the same discrepancy between FRDM and Skyrme functionals in the case of Z=120 and Z=126 isotopes. These observations indicate that the FRDM fails to describe correctly the $\alpha$-decay properties in these isotopes

To sum up, from the all investigated quantities, it is proven that the effect of the tensor force is significantly important. This can respond positively to the warning that summarized in a recent review excerpt from \cite{Otsuka2020}: "Regarding open problems with Skyrme-based approaches, we quote the comment "the currently used central and spin-orbit parts of the Skyrme energy density functional are not flexible enough to allow for the presence of large tensor terms" from Lesinski
et al. (2007), and another remark, "studies of tensor terms are extended to the case with deformations for future construction of improved density functionals" from Bender et al. (2009). It is another open question as to what extent observed states are of a single-particle nature."\\
Note that Lesinski et al. (2007) and Bender et al. (2009) are nothing but \cite{Lesinski} and \cite{Bender}, respectively.

\section{Conclusion}\label{Sec5}
In this paper, the impact of the tensor force on the evolution of neutron shell closures has been systematically investigated for even-even super-heavy nuclei, which are assumed to be the next magic nuclei beyond Pb, with Z=114, 120 and 126. This study is performed by means of ten different Skyrme functionals in two cases with and without tensor force in the framework of the Hartree-Fock+BCS approximation. It is found that the tensor force modifies significantly several physical observables of the nuclei in the ground state, especially at shell closure N=184, such as: Two-neutron separation energies, two-neutron shell gap, neutron pairing gap, root mean-square radii, neutron single particle energies, Q$_{\alpha}$-energy and $\alpha$-decay half-live.
It is also shown that the impact of the tensor force on these observables is more pronounced for Skyrme parametrizations which have a large isoscalar tensor coupling constant $C_0^J$. For instance, T22, T24, T42 and SLy5 with  lower $C_0^J$ are less influenced by the tensor force compared to T26, T44 and T62 with medium $C_0^J$, whereas the influence is intense for T46, T64 and T66 with large $C_0^J$. Furthermore, it is found that different Skyrme sets with tensor force lead to a same evolution trend, compared to those without tensor force, not only for the single-particle energies but also for all several studied observables. 
In addition, the comparison of our DFTs results with mic-mac FRDM predictions does not always show significant agreement. The peak observed in $S_{2n}$, at N=184 for Z=126, the negative values in $\delta_{2n}(Z,N)$ at (Z=114, N=188), (Z=120 , N=186) and (Z=126, N=182) which is contradictory with the definition of this entity which should be positive and the unexpected trend seen in the $Q_{\alpha}$-energy and $\alpha$-decay half-lives demonstrate that the widely accepted theoretical predictions made by the macro-microscopic Finite Range Droplet Model (FRDM) fail to describe the neutron shell closure N=184. This can be explained by the lack of self-consistency effect in the micro-macroscopic models with respect to DFT framework.
 Finally, it is worth mentioning  that in this work we only focused on investigating the tensor force impact on the neutron shell closure of super-heavy elements. We think that the proton shell closure of this elements could also be influenced by the tensor force. The work at this direction is in progress.

\section*{Aknowledgement} This research was supported through computational resources of HPC-MARWAN (www.marwan.ma/hpc) provided by the National Center for Scientific and Technical Research (CNRST), Rabat, Morocco 


\end{document}